\documentclass[11pt,a4paper]{article}
\usepackage{jheppub}
\usepackage{amsmath,amssymb,amsfonts,bbold}
\usepackage{graphicx,bm}
\usepackage{verbatim}
\usepackage{diagbox}
\usepackage{xcolor}
\usepackage{hyperref}

\definecolor{darkgreen}{RGB}{30,98,41}

\definecolor{bluu}{rgb}{0.000111,0.001760,0.998218}
\definecolor{orr}{rgb}{0.990416,0.500269,0.032866}
\definecolor{grr}{rgb}{0.131345,0.999677,0.023624}

\title{Master equations and stability of Einstein-Maxwell-scalar black holes}
\author[a]{Aron Jansen}
\author[b]{Andrzej Rostworowski}
\author[b]{Mieszko Rutkowski}
\affiliation[a]{Departement de F\'{i}sica Quantica i Astrof\'{i}sica, Institut de Ci\'{e}ncies del Cosmos, Universitat de Barcelona, Mart\'{i} i Franqu\'{e}s 1, E-08028 Barcelona, Spain} 
\affiliation[b]{M. Smoluchowski Institute of Physics, Jagiellonian University, 30-348 Krakow, Poland}
\emailAdd{a.p.jansen@icc.ub.edu}
\emailAdd{arostwor@th.if.uj.edu.pl}
\date{\today}

\abstract{
We derive master equations for linear perturbations in Einstein-Maxwell scalar theory, 
for any spacetime dimension $D$ and any background with a maximally symmetric $n = (D - 2)$-dimensional spatial component.
This is done by expressing all fluctuations analytically in terms of several master scalars.
The resulting master equations are Klein-Gordon equations, with non-derivative couplings given by a potential matrix of size 3, 2 and 1 for the scalar, vector and tensor sectors respectively.
Furthermore, these potential matrices turn out to be symmetric, and positivity of the eigenvalues is sufficient (though not necessary) for linear stability of the background under consideration.
In general these equations cannot be fully decoupled, only in specific cases such as Reissner-Nordstr\"{o}m, where we reproduce the Kodama-Ishibashi master equations.
Finally we use this to prove stability in the vector sector of the GMGHS black hole and of Einstein-scalar theories in general.
}

\begin{document}
\maketitle

\section{Introduction}
General Relativity admits a wide variety of Black Hole solutions, especially when coupled to matter.
For any such black hole, or black brane, one of the central questions is how it behaves when perturbed, and in particular whether or not it is a (linearly) stable solution.

To answer this question one has to solve the linearized perturbation equations.
This can be a messy task, because the metric, and the matter fields, have many components which all need to be fluctuated, and which may all couple.

The field of black holes perturbations was born with the seminal Regge-Wheeler \cite{Regge:1957td} paper, where the stability of the Schwarzschild solution under \textit{axial} linear perturbations was analysed. 
This study was extended to \textit{polar} perturbations by Zerilli \cite{Zerilli:1970se,Zerilli:1971wd} and given a new perspective by Moncrief \cite{Moncrief:1974am}. 
Later, it was extended to the Schwarzschild-de Sitter \cite{Mellor:1989ac} and Schwarzschild-anti-de Sitter \cite{Cardoso:2001bb} backgrounds. 
Finally, the problem of linear fluctuations was treated in full generality by Kodama\&Ishibashi (KI). 
In the outstanding work~\cite{Kodama:2003jz} the problem was generalized for perturbations of general maximally-symmetric black holes i.e. to arbitrary spacetime dimension, 
cosmological constant and to any of the three maximally symmetric horizon topologies (spherical, planar and hyperbolic), and then extended to include Maxwell field (electro-vacuum) in~\cite{Kodama:2003kk}. 
These master equations have also been generalised to Gauss-Bonnet gravity in \cite{Dotti:2005sq,Gleiser:2005ra} and to Lovelock gravity in \cite{Takahashi:2010ye}.
\\
The key result of black hole perturbation theory is that the general perturbation can be given in terms of only few scalar functions satisfying scalar wave equation with some potentials. 
In particular, in the case of maximally symmetric black holes with an electric charge \cite{Kodama:2003kk} it was shown that the full problem of linear fluctuations can be reduced to solving 5 fully decoupled scalar wave equations with potentials, the so called \textit{master scalar wave equations}. 
These are equations on the \textit{master scalars}, in which all the fluctuations are expressed in a fully analytic way.
\footnote{More precisely, in higher dimensions, $D>4$, the master scalars in vector and tensor sectors come in a number of copies corresponding to different polarizations of gravitational waves in these sectors.}. 
Moreover, this structure can be extended beyond linear approximation and we find it really remarkable that in metric perturbation approach to Einstein equations, solving a full set of perturbation Einstein equations (at any perturbation order) can be reduced to the problem of solving a couple of scalar wave equations~\cite{Rostworowski:2017ruj, Rutkowski:2019tag}.

In this work we generalize Kodama\&Ishibashi (KI) results~\cite{Kodama:2003kk} to theories which in addition to a charge have a scalar field, with an arbitrary potential and an arbitrary potential coupling to the gauge field. 
This covers a broad class of actions, which have been heavily studied in the context of holography \cite{Maldacena:1997re}. 
The simplest examples are Einstein-scalar theories in anti de Sitter spacetimes, with the physics depending heavily on the choice of potential and the background solutions typically being numerical, see e.g.~\cite{Gubser:2008ny,Gursoy:2008za,Janik:2016btb}.
Including the gauge field some analytic examples are the GMGHS (Gibbons, Maeda, Garfinkle, Horowitz, Strominger) black hole~\cite{Gibbons:1987ps,PhysRevD.43.3140} of which we will analyze the stability in section \ref{sec:strominger}, and the asymptotically Lifshitz black brane of~\cite{Tarrio:2011de}.
 
Although our results are an extension of KI, our derivation (initiated in \cite{Rostworowski:2017ruj,Rostworowski:2019iig}) is slightly different.
Instead of making manipulation with linearized Einstein equations (consisting mainly in taking different linear combination  of these equations and their derivatives to arrive at master scalar equations), 
we take the structure of the outcome of previous work as the initial input for our procedure: we make an \textit{ansatz} that all gauge invariant characteristics of fluctuation (see below for their definition) are given in terms of linear combinations of master scalars and their derivatives, 
where the master scalar themselves satisfy scalar wave equations coupled with interaction potentials. 
As the final results, we express all the perturbations analytically in terms of the three sets of master scalars, one for each helicity $h$. 
We find such an \textit{ansatz approach} to solve for the fluctuations to be a very robust technique: interestingly it works also for time-dependent backgrounds, for example in the cosmological perturbations context \cite{Rostworowski:2017ruj,Rostworowski:2019iig}). 
The main advantage is that once we decide on the correct form of the ansatz (i.e. the highest order of derivatives of master scalars in the linear combinations for gauge invariants and/or the form of the couplings between master scalars in master scalar wave equations), 
finding the function coefficients of these linear combinations and the actual form of the coupling potentials is an purely algorithmic task (although rather unthinkable to achieve in the pre-computer algebra packages era).

We express all the perturbations analytically into three sets of master scalars, one for each helicity $h$. 
\\
Each set of master scalars satisfies a coupled master equation of the form,
\begin{equation}\label{eq:masterequation}
\square \Phi^{(h)}_s - W{}^{(h)}_{s,s^\prime}(r) \Phi^{(h)}_{s^\prime} = 0 \, ,
\end{equation}
where $\square$ stands for the wave operator on the background metric (see eq.(2.2) below) and  the potential matrix $W$ couples the different master scalars with the same helicity, which are labeled by their spin.
The components of the perturbations are first expressed into gauge invariant combinations (Eq. \ref{eq:gaugeInvariants}), which are then expressed in terms of these master scalars (Eqs. \ref{eq:InvsToMasterH0}, \ref{eq:InvsToMasterH1} and appendix \ref{app:scalarsector}).
The potentials are given in Eqs. \ref{eq:potTensor}, \ref{eq:potVector}, \ref{eq:potScalar1} and \ref{eq:potScalar2}.
In section \ref{sec:fulldecoupling} we discuss when these still coupled master equations can be further decoupled into single equations.
Then in section \ref{sec:stability} we discuss a sufficient criterion for linear stability and apply it to several specific cases.

The master equations we derive here are made available in a Mathematica notebook, along with a check of their correctness, at~\cite{githublink}.

\section{Setup}
We consider the class of Einstein-Maxwell-scalar theories described by the following action,
\begin{equation}\label{eq:action}
S = \int d^{n+2} x \sqrt{-g} \left(R - 2 \Lambda - \eta (\partial \phi)^2 - \frac{1}{4} Z(\phi) F^2 - V(\phi) \right) \, ,
\end{equation}
where $R$ is the Ricci scalar, $\Lambda$ is a cosmological constant, $\eta$ is an arbitrary normalization factor for the scalar field $\phi$\footnote{This can be absorbed into $\phi$ but we keep it explicit to make it easier to substitute a particular model.}, $F = d A$ is the field strength and $V$ and $Z$ are two arbitrary functions of the scalar field, with $V(0) = 0$.

Any time-independent $n+2$ dimensional solution with a maximally symmetric $n$-dimensional spatial part can be written as\footnote{Note that we could set either $\zeta = 1$ or $S = r$ by a gauge transformation, but we choose to keep it in this more general form.}
\begin{equation}\begin{split}\label{eq:metricAnsatz}
ds^2 &= - f(r) dt^2 + \frac{\zeta(r)^2}{f(r)} dr^2 + S(r)^2 dX_{(n,K)}^2 \, , \\
A &= a(r) dt \, . 
\end{split}\end{equation}

Here $dX_{(n,K)}^2$ is one of the three maximally symmetric $n$-dimensional spaces,
\begin{equation}\label{eq:spatialAnsatz}
dX_{(n,K)}^2 =  
\begin{cases}
d x_1^2 + ... + d x_{n}^2, &K =\,\, \,0 \, , \,\,\,\,\, \text{planar}\, \\
d \Omega_{(n)}, &K = +1 \, , \ \text{spherical}\,  \\
d H_{(n)}, &K = -1 \, , \ \text{hyperbolic} \, 
\end{cases} \, .
\end{equation}

Here we note that while Kodama\&Ishibashi approach \cite{Kodama:2003jz,Kodama:2003kk} is coordinate independent we prefer to work with the fixed Fefferman-Graham, or Schwarzschild-like, coordinate system (\ref{eq:metricAnsatz}). However, since our final result, eq. (\ref{eq:masterequation}) is a scalar equation it can be easily expressed in any coordinate system; similarly the rules that express gauge invariant quantities in terms of master scalars can be easily transformed (see Appendix \ref{app:transformations}).

In order to avoid cluttering the presentation, we leave any complications relating to spherical or hyperbolic symmetry to Appendix \ref{app:spherical}, focussing here on the planar case.
In the results presented we do show the most general expressions, where the dependence on the topology shows up only through the parameter $K$ defined above and the eigenvalues $-k^2$ of the corresponding Laplace operator.

The equations of motion following from Eq. (\ref{eq:action}) lead to the following equations for the background\footnote{Here and in everything that follows, we extrapolate the dimensional dependence from our calculations at $n = 2, ..., 9$.}:
\begin{equation}\label{eq:bg}\begin{split}
\phi'' &=\phi ' \left(\frac{\zeta '}{\zeta }-n \frac{ S'}{S}\right)-\frac{a'^2 Z'+4 \eta  f' \phi '-2 \zeta ^2 V'}{4 \eta f } \, , \\
a'' &=a' \left(\frac{\zeta '}{\zeta }-n \frac{S'}{S}-\frac{Z' \phi '}{Z}\right) \, , \\
S'' &=\frac{\zeta ' S'}{\zeta }-\frac{\eta}{n}  S \phi'^2 \, , \\
0 &= S^2 \left(2\eta f \phi'^2-Z a'^2\right)-2n S f' S'-2 n (n-1) f S'^2+2 \zeta ^2 \left(n(n-1) K-S^2 (V + \Lambda) \right) \, , \\
f'' &=Z a'^2+\frac{f' \zeta '}{\zeta }-(n-2)\frac{f' S'}{S}-\frac{2(n-1)}{S^2} \left(\zeta ^2 K-f S'^2\right)-\frac{2\eta}{n} f \phi'^2 \, ,
\end{split}\end{equation}
where in a slight abuse of notation, primes indicate radial derivatives except when acting on $V$ or $Z$, where they indicate a derivative with respect to $\phi$.

We further note that although we will work with the Fefferman-Graham coordinates Eq. (\ref{eq:metricAnsatz}) here, the final potentials in the master equations Eq. (\ref{eq:masterequation}) will be exactly equal in the Eddington-Finkelstein coordinates parametrized as,
\begin{equation}\label{eq:EFansatz}
ds^2 = - f(r) dt^2 + 2 \zeta(r) dt dr + S(r)^2 dX_{(n,K)}^2 \, , 
\end{equation}
Any differences between the two will be shown in Appendix \ref{app:transformations}.

We will not specialize to any specific background, but consider any background that satisfies these equations.

In the following section we will perturb this general solution and derive the master equations that describe these perturbations.

\section{Master Equations}

Perturbing the background solution to first order, we have to perturb all the fields: the metric, the gauge field and the scalar.
Because of the maximal symmetry of the spatial part of the background, we can express the spatial dependence of these fluctuations using the eigenfunctions of the Laplacian of the $n$-dimensional maximally symmetric space, which in the present planar case are just plane waves, giving the following perturbations:

\begin{equation}\begin{split}\label{eq:metricfluct}
\delta g_{\mu\nu} &= h_{\mu\nu}(t,r) e^{i k x} \, , \\
\delta A_\mu &=  a_\mu(t,r) e^{i k x} \, , \\
\delta \phi &= \varphi(t,r) e^{i k x} \, ,
\end{split}\end{equation}
where we've chosen the plane waves to propagate along the first spatial coordinate in $X = \left( x \equiv x_{(1)}, y \equiv x_{(2)}, ..., z \equiv x_{(n)} \right)$.

The derivation of the master equations takes the following steps:
\begin{enumerate}
\item Organise the fluctuations into three different sectors or channels, according to their transformations under the little group. \, \\
\item Rewrite the the fluctuations into gauge-invariant combinations (or equivalently choose a gauge that is fixed uniquely).
\item Rewrite those gauge-invariant combinations as linear combinations of master scalars and their derivatives, where the master scalars themselves satisfy Eq. (\ref{eq:masterequation}).
\end{enumerate}

We shall now discuss each step in turn.

\subsection{Sectors}

The perturbations Eq. (\ref{eq:metricfluct}) naturally decouple into three sets of equations, as summarized in Table (\ref{tab:sectors}).
The sectors are classified by their helicity, 0, 1 or 2, or whether the fluctuation transforms as a scalar, vector or tensor once the momentum is fixed.
Various different names are used in the literature for these sectors, we will stick to scalar, vector and tensor since these seem to be the most natural and context-independent.

In Table (\ref{tab:sectors}) we summarize how the different components fall into the three sectors.
We adopt a convention where indices $i,j$ take values from $(t,r,x)$ and indices $\alpha \neq \beta$ take values from $(x_{2} \equiv y, ..., x_{n}\equiv z)$.

\begin{table}[htp]
\begin{center}
\begin{tabular}{c||c|c|c|c|c||}
\diagbox{h}{s} & $h_{\mu\nu}$ 																		& $a_\mu$ 					& $\varphi$ 				& copies $\times N_\text{coupled}$	& names										\\ \hline\hline
0	& \shortstack{$h_{ij}\, \quad h$ \\ $6\,  \quad \,\,\,1$ } 												& \shortstack{$a_i$\\$3$} 			& \shortstack{$\varphi$\\1} 	& $1 \times 11$  				& \shortstack{\underline{scalar}\\ parity-even, polar \\sound}			\\ \hline
1	& \shortstack{$h_{i\alpha}$ \\ $(n-1)\times 3$}														& \shortstack{$a_\alpha$\\$n-1$} 	& $-$ 					& $(n-1) \times 4$ 				& \shortstack{\underline{vector}\\ parity-odd, axial \\ shear}			\\ \hline
2	& \shortstack{$h_{\alpha,\beta} \, \quad h_{\alpha \alpha} - h_{\beta\beta}$\\ $\frac{1}{2} (n-1)(n-2)\, \quad n-2$} 	& $-$ 						& $-$ 					& $\frac{1}{2}(n+1)(n-2)  \times 1$	& \shortstack{\underline{tensor}\\ scalar}				\\ \hline
total	& $\frac{1}{2} (n+2)(n+3)$ 																	& $n+2$ 						& 1 						& $\frac{1}{2} (n+3)(n+4)$  		&											\\ \hline\hline
\end{tabular}
\end{center}
\label{tab:sectors}
\caption{Decoupling of perturbations into sectors. Under each perturbation we note the number of components involved. In the rightmost column we list some names that are common in the literature for these sectors, here we use the underlined ones.}
\end{table}%

Each of the $1 + (n-1) + (1/2)(n+1)(n-2) = (1/2) n (n+1)-1$ copies in the fourth column will include one gravitational master scalar. 
This number is equal to the number of graviton polarisations, which can be counted as a symmetric $n$ by $n$ matrix, subtracting the trace.

The scalar sector is the most complicated one, since it receives contributions from every field. In particular the scalar field itself of course falls into this sector. So do the components of the gauge field and metric with $i$ indices, and finally the trace of the spatial metric perturbations, $h$. In the table we also list the number of components each of these has. For the scalar sector there are $11$ fluctuations in total, which all couple to each other. 
In the next subsection we will show that the $11$ coupled PDEs for fluctuations in this sector are in fact $11$ equations for $7$ gauge invariant characteristics of scalar perturbations.

The vector sector consists of those fluctuations with one $\alpha$ index. This can be the gauge field, or the metric where the other index is an $i$. Together these give $4$ components, times $(n-1)$ for the number of values that $\alpha$ can take. 
These are not all coupled though, this sector further decouples into $(n-1)$ identical copies of sets of 4 coupled equations, one for each value of $\alpha$.
In the next subsection we will show that each copy of the $4$ coupled PDEs for fluctuations in this sector is in fact a copy of $4$ equations for $3$ gauge invariant characteristics of vector perturbations.

Finally the tensor sector can only have contributions from the metric, and consists of those metric fluctuations with two distinct indices $\alpha,\beta$, of which there are $\frac{1}{2} (n-1)(n-2)$, and differences of diagonal components, of which there are $n-2$ linearly independent ones.
This sector is particularly simple, since none of these couple to each other. Thus it falls into $(1/2)(n+1)(n-2)$ decoupled equations. 
All components in this sector are in fact gauge invariant - see the next subsection.

Since the tensor sector equations are all identical we can consider only one tensor perturbation, which we will take to be $h_{y z}$.
Furthermore since the vector sector consists of identical sets of coupled equations for each value of $\alpha$ we can also consider only one copy of those, which we shall take along the $z$ direction, so we perturb $h_{t z}, h_{r z}, h_{x z}$ and $a_z$.
In the scalar sector we need all 11 perturbations, but instead of $h_{x x}$ and $h$ we use different linear combinations.

The perturbations we take are:
\begin{equation}\label{eq:components}
\begin{split}
\delta g_{\mu\nu} &= 
\begin{pmatrix}
h_{t t} 			&1/2 h_{t r} 		& i k \, h_{t x} 	& 0 					& ... 		& 0 			& h_{t z} 				\\
1/2 h_{t r} 			& h_{r r}			& i k \, h_{r x} 	& 0 					& ... 		& 0 			& h_{r z} 				\\
i k \, h_{t x} 		& i k \, h_{r x}		& h_{x x} 	& 0 					& ... 		& 0 			& i k h_{x z} 			\\
0				& 0				& 0			& h_{y y}	& 0		& 0			& h_{y z}				\\
\vdots			& \vdots 			& \vdots		& 0					& \ddots	& 0			& 0					\\
0				& 0				& 0			& 0					& 0		& \,\,\,\ddots	& 0					\\
h_{t z}			& h_{r z}			& i k h_{x z}	& h_{y z}				& 0		& 0			& h_{z z}	\\
\end{pmatrix}  e^{i k x} \, , \\
\delta A_{\mu} &= 
\begin{pmatrix}
a_t\,,				& a_r\,, 			& i k a_x\,,	 	& 0\,, 				& ...\,, 	& 0\,, 		& a_z\,				\\
\end{pmatrix}  e^{i k x} \, , \\
\delta\phi &= \varphi e^{i k x} \, ,
\end{split}\end{equation}
where each function now depends on $(t,r)$, and we further rewrite:

\begin{equation*}\begin{split}
h_{x x} &= \frac{1}{n}\left( h_+ - (n-1) k^2 h_- \right) \, , \\ 
h_{yy} =\dots =h_{zz}&= \frac{1}{n}\left( h_+  + k^2 h_- \right) \, .
\end{split}\end{equation*}

This particular convention comes from the decomposition into scalar, vector and tensor components for general maximally symmetric topologies that we do in Appendix \ref{app:spherical}, taking the planar case.

\subsection{Gauge Invariant Fluctuations}\label{sec:gaugeinv}

We can now use gauge transformations to simplify this further.

We do an infinitesimal coordinate transformation $x^\mu \rightarrow x^\mu + \xi^\mu$ and an infinitesimal gauge transformation $A_\mu \rightarrow A_\mu + \nabla_\mu \lambda$, 
where $\xi^\mu$ and $\lambda$ are arbitrary functions of $(t,r)$.
If we keep the background fields invariant, the perturbations have to transform as,
\begin{equation}\begin{split}
\delta g_{\mu\nu} &\rightarrow \delta g_{\mu\nu} - \nabla_\mu \xi_\nu - \nabla_\nu \xi_\mu \, , \\
\delta A_\mu &\rightarrow \delta A_\mu + \nabla_\mu \lambda - \xi^\nu \nabla_\nu A_\mu - A_\nu \nabla_\mu \xi^\nu \, , \\
\delta \phi &\rightarrow \delta \phi - \xi^\nu \nabla_\nu \phi \, .
\end{split}\end{equation}

Now we further decompose the fluctuations into gauge-independent and gauge-dependent ones.
We find the following set of gauge-independent combinations:
\begin{equation}\label{eq:gaugeInvariants}\begin{split}
\text{helicity 2: }\\
\mathfrak{h}_{y z} &\equiv h_{y z} \, , \\
\text{helicity 1: } \\
\mathfrak{h}_{t z} &\equiv h_{t z} - \partial_t h_{x z}\, , \\
\mathfrak{h}_{r z} &\equiv h_{r z} - \partial_r h_{x z} + 2 \frac{S'}{S} h_{x z}\, , \\
\mathfrak{a}_{z} &\equiv a_z \, , \\
\text{helicity 0: } \\
\mathfrak{h}_{t t} &\equiv h_{t t} - 2 \partial_t h_{t x} + \partial_t^2 h_- + \frac{f'}{2 n S S'} (h_+ + k^2 h_-)\, , \\
\mathfrak{h}_{t r} &\equiv h_{t r} - 2 \partial_r h_{t x} + \partial_t \partial_r h_- + 2 \frac{f'}{f} h_{t x} - \frac{f'}{f} \partial_t h_- -  \frac{\zeta^2}{n f S S'} \partial_t (h_+ + k^2 h_-) \, , \\
\mathfrak{h}_{r r} &\equiv h_{r r} - \frac{\zeta^2}{n f S' S}\partial_r (h_+ + k^2 h_-) + \left(\frac{\zeta^2}{2 n f^2 S^2 S'} (S f' + 2 f S') - \eta \frac{\zeta^2}{n^2 f S'^2}\phi'^2\right) (h_+ + k^2 h_-) \, , \\
\mathfrak{h}_{r x} &\equiv h_{r x} - \frac{1}{2} \partial_r h_- + \frac{S'}{S} h_- -  \frac{\zeta^2}{2 n f S S'} (h_+ + k^2 h_-) \, , \\
\mathfrak{a}_{t} &\equiv a_{t} - \partial_t a_x -  \frac{a'}{2 n S S'} (h_+ + k^2 h_-) \, , \\
\mathfrak{a}_{r} &\equiv a_{r} - \partial_r a_x + \frac{a'}{2 f} \partial_t h_- -  \frac{a'}{f} h_{t x}\, , \\
\text{\boldmath$\varphi$}&\equiv \varphi -  \frac{\phi'}{2 n S S'} (h_+ + k^2 h_-) \, , \\
\end{split}\end{equation}
Note here that although $K$ does not appear in these expressions, these \textit{are} the correct expressions for any $K$. Although these expressions are independent of $K$, the definition of the components through Eq. (\ref{eq:components}) does need to be modified, see Appendix \ref{app:spherical}.

Note the structure of these definitions. The gauge invariants are formed by some subset of the components, ``dressed'' with the other components and their derivatives to make them gauge invariant.
In the tensor sector this is trivial. 
If we demand this structure and that the coefficients are known algebraically in terms of the background, this choice of gauge invariants is unique in the vector sector.
In the scalar sector there is however another choice. The one we have chosen is the Detweiler gauge \cite{Chen:2016plo, Corrigan:2018svy}. Instead one could have chosen the Regge-Wheeler gauge, where $h_+$ is taken as the basis for a gauge invariant instead of $h_{r x}$.
We find the Detweiler gauge simpler to work with, however we note that since the master equations are gauge invariant, it actually does not matter for the final potentials. 
Intermediate results in the Regge-Wheeler gauge are discussed in Appendix \ref{app:transformations}.

When these gauge invariants are substituted into the perturbation equations, the remaining non-gauge-invariant ``dressing'' components ($h_{x z}$ in the vector sector and $h_{t x}$, $h_\pm$ and $a_x$ in the scalar sector) automatically drop out.
Instead of using the gauge invariant components one may use the gauge freedom to set these components to zero, fixing the gauge.
We stress that fixing the gauge, to Regge-Wheeler or Detweiler gauge, is completely equivalent to working with gauge-invariant variables.

This step, and the next, are summarized in Table \ref{tab:gaugeComponents}.

\begin{table}[htp]
\begin{center}
\begin{tabular}{c||c|c|c||}
\diagbox{h}{s} & $h_{\mu\nu}$ 																			& $a_\mu$ 											& $\varphi$ 							\\ \hline\hline
0	& \shortstack{$\mathfrak{h}_{t t} ,\,  \mathfrak{h}_{t r} ,\, \mathfrak{h}_{r r} ,\, \mathfrak{h}_{r x}$ \\ $h_{t x}\,(\xi_t) ,\, h_+\, (\xi_r) ,\, h_-\, (\xi_x)$ \\ $\Phi^{(0)}_2$ } 	& \shortstack{$\mathfrak{a}_t ,\,  \mathfrak{a}_r$ \\ $a_x\,(\lambda)$ \\ $\Phi^{(0)}_1$ } 	& \shortstack{$\text{\boldmath$\varphi$}$ \\ - \\ $\Phi^{(0)}_0$ }	\\ \hline
1	& \shortstack{$\mathfrak{h}_{t z} ,\, \mathfrak{h}_{r z}$ \\ $h_{x z}\, (\xi_z)$  \\ $\Phi^{(1)}_2$ } 										& \shortstack{$\mathfrak{a}_z$ \\ - \\ $\Phi^{(1)}_1$} 						& -									\\ \hline
2	& \shortstack{$\mathfrak{h}_{y z}$ \\ - \\ $\Phi^{(2)}_2$ } 															& -	 												& -	 								\\ \hline\hline
\end{tabular}
\end{center}
\label{tab:gaugeComponents}
\caption{Decoupling of sectors into gauge-independent components. For each sector and each field we list first the gauge-invariant components, then in the line below the gauge-dependent ones and behind them in brackets the gauge parameter that can be used to set it to zero.
The bottom line is the master scalar of that field in that sector.}
\end{table}%

\subsection{Master Equations}

The former two steps, the decoupling of independent sectors and the decoupling of gauge-dependent modes, are quite standard and technically simple, and it is no surprise that this can be done.
The final step from the gauge invariant fluctuations to the master scalars is technically more difficult, and here it is also not clear why the equations can be written in the simple form that we will see.

Conceptually however this step is also very simple.
We assume that we can express all the fluctuations at a given spin and helicity into a single so-called ``master scalar'', which satisfies a Klein-Gordon equation with a certain potential.
We make an ansatz for the coefficients relating the gauge-invariant components to the master scalars, and for the potentials.
Then we insert this ansatz into the perturbation equations and try to find a solution.

Around a vacuum solution, where the gauge field and scalar are zero in the background and only consist of the fluctuations, the different spins also decouple.
So for a given helicity and spin $(h,s)$ we can express all the fluctuations in terms of a single master scalar $\Phi^{(h)}_s$ that satisfies a Klein-Gordon equation with a potential $W^{(h)}_{s,s}$.

The coefficients and potentials in the ansatz are found by plugging the ansatz into the perturbation equations, and using the background equations (\ref{eq:bg}) and the master equations (\ref{eq:masterequation}) (which involve the as yet unknown potentials) to simplify.
In each equation every coefficient of the master scalar and its derivatives must individually vanish, provided the master equations have been imposed.
This system of equations is such that for a given helicity and spin one has to solve a single simple first order ODE, that can be solved analytically. This gives one integration constant for each spin and helicity, corresponding to an arbitrary normalization of the master scalar.
The rest of the equations are algebraic and can easily be solved, although in practice it can be rather difficult even with Mathematica.

What changes when there is a gauge field and/or scalar field in the background, is that now the master scalars in a given sector couple through the non-derivative interaction potentials $W^{(h)}_{s,s^\prime}$.
Remarkably, and non-trivially, these interaction potentials can be made symmetric through a choice of the aforementioned integration constants. 
This leaves one free integration constant per sector, that does not affect the potentials but just scales all the master equations by the same constant.
In order to get the master equations in this form, the gauge-invariants also have to receive contributions from the other master scalars at different spins.

We will now look at the results of this procedure sector by sector.

\subsubsection{Tensor sector}

In the tensor sector, only present for $n > 2$, nothing changes with respect to the vacuum case.
The fluctuation is proportional to the master scalar as,
\begin{equation}\label{eq:InvsToMasterH0}
\mathfrak{h}_{y z} \equiv S^2 \Phi^{(2)}_2 \, ,
\end{equation}
and the master scalar satisfies a free Klein-Gordon equation,
\begin{equation}\label{eq:potTensor}
W{}^{(2)}(r) = \frac{k^2}{S^2} \, .
\end{equation}
Note that this single term comes simply from the Laplacian acting on a scalar eigenfunction, meaning that tensor modes satisfy a free, massless scalar field equation.

\subsubsection{Vector sector}

The vector sector, which is present only for $n > 1$, is still quite simple and similar to the vacuum case, in that there is no mixing between different fields at the level of the master scalars,
\begin{equation}\label{eq:InvsToMasterH1}\begin{split}
\mathfrak{h}_{t z} &\equiv \frac{n}{k \tilde{k}} \frac{f S S'}{\zeta} \Phi^{(1)}_2 +\frac{1}{k \tilde{k}} \frac{f S^2}{\zeta} \partial_r \Phi^{(1)}_2 \, , \\
\mathfrak{h}_{r z} &\equiv \frac{1}{k \tilde{k}} \frac{\zeta S^2}{f} \partial_t \Phi^{(1)}_2 \, , \\
\mathfrak{a}_{z} &\equiv \frac{1}{k} \frac{S}{\sqrt{Z}}\Phi^{(1)}_1 \, ,
\end{split}\end{equation}
where we have defined $\tilde{k} \equiv \sqrt{k^2 - n K}$ and we have chosen an overall normalization of the master scalars to reflect the singularity of the zero momentum, or in spherical setting $l = 1$, limits, see appendix \ref{app:specialCasesVector}.

However as mentioned we get two equations which are coupled through the potential matrix,
\begin{equation}
W^{(1)} = 
\begin{pmatrix}
W{}^{(1)}_{1,1} 			& W{}^{(1)}_{1,2} 			\\
W{}^{(1)}_{1,2} 			& W{}^{(1)}_{2,2}			
\end{pmatrix} \, ,
\end{equation}
where
\begin{equation}\label{eq:potVector}\begin{split}
W{}^{(1)}_{1,1}(r) &= \frac{k^2}{S^2} - \frac{f' S'}{\zeta ^2 S}  +(n-2) \left(\frac{K}{S^2}-\frac{f S'^2}{\zeta ^2 S^2}\right)     +    \frac{Z a'^2}{\zeta ^2}    +   \frac{f \eta  \phi'^2}{n \zeta ^2} -   \frac{1}{8\eta \zeta^2 } \frac{Z'}{Z}  \mathcal{V}   \\
   & -   \frac{Z'^2}{Z^2} \frac{f \phi'^2}{4 \zeta ^2}  - \frac{Z'}{Z}\frac{f S' \phi '}{\zeta ^2S}   +   \frac{f \phi'^2 Z''}{2 \zeta ^2 Z}\, , \\
W{}^{(1)}_{1,2}(r) &= - \sqrt{k^2 - n K} \frac{\sqrt{Z} a' }{\zeta  S} \, , \\
W{}^{(1)}_{2,2}(r) &= \frac{k^2}{S^2} -n \left(\frac{f' S'}{\zeta ^2 S}-\frac{f S'^2}{\zeta ^2 S^2}+\frac{K}{S^2}\right) + \eta \frac{f  \phi'^2}{\zeta ^2}\, ,
\end{split}\end{equation}
and
\begin{equation}
\mathcal{V}(r) = - 2 \zeta^2  V' +  a'^2 Z' \, .
\end{equation}
\\
Note that as expected, when the background gauge field vanishes, the equations decouple.
Also note that the factor $\sqrt{k^2 - n K}$ in the interaction potential. 
In the planar case this becomes simply $k$, and the equations decouple in the zero momentum limit.
In the spherical case this factor becomes equal to $\sqrt{(l-1)(l+n)}$, so that the equations again decouple if $l = 1$.
This makes sense too, because at $l=1$ there are no dynamical degrees of freedom in the metric, only in the gauge field.

\subsubsection{Scalar sector}

Here it gets significantly more complicated, and we present the gauge invariants in Appendix \ref{app:scalarsector}.
These are expressed in terms of three master scalars $ \Phi^{(0)}_2,  \Phi^{(0)}_1$ and $ \Phi^{(0)}_0$, which in vacuum would correspond to the gravitational, gauge field and scalar fluctuations respectively.
In the general case however, both the metric and the gauge field fluctuations receive contributions from all three master scalars, while the scalar field gets a contribution from the gravitational master scalar in addition to its own.

These three master scalars again satisfy the coupled Klein-Gordon equations \ref{eq:masterequation} with the potential matrix,
\begin{equation}
W^{(0)} = 
\begin{pmatrix}
W{}^{(0)}_{0,0} 			& W{}^{(0)}_{0,1} 			& W{}^{(0)}_{0,2} 			\\
W{}^{(0)}_{0,1} 			& W{}^{(0)}_{1,1} 			& W{}^{(0)}_{1,2} 			\\
W{}^{(0)}_{0,2} 			& W{}^{(0)}_{1,2} 			& W{}^{(0)}_{2,2} 			\\			
\end{pmatrix} \, .
\end{equation}

The three diagonal potentials are,
\begin{equation}\begin{split}\label{eq:potScalar1}
W{}^{(0)}_{0,0}(r) &= \frac{k^2}{S^2} + \frac{\phi' }{ \mathcal{D}^2 \zeta ^2} \Biggl( 
\frac{\zeta^2 k^2}{n S'} \mathcal{A} + \mathcal{F} \mathcal{D} S \mathcal{V}  + 2 \eta f \phi' \left( \mathcal{F} \mathcal{P}+4 \zeta ^4 k^2 \left(k^2-(n-1)K \right)\right) + \\
& 2 \eta \mathcal{F} \mathcal{D}  \phi' \left(S f'+  (n-2) f  S'  \right)  \Biggr) -\frac{1}{4 \eta  \zeta ^2} \left(\mathcal{V}'- 2 a'^2  Z'^2/Z \right)  \, , \\
W{}^{(0)}_{1,1}(r) &= \frac{k^2}{S^2} + \frac{Z a'^2}{\mathcal{D}^2 \zeta ^2} \Biggl(n^2 S'^2 \mathcal{F} \left(S f'-2  (n-1) f S'\right) + 2 f n^2 S^2 Z a'^2 S'^2 + \\
&4 f \zeta ^2 \left(n S'^2 \left((2n-3)k^2 - n(n-1) K \right)+k^2 \eta  S^2 \phi'^2\right)+4 \zeta ^4 k^4\Biggr) + \\
&\frac{1}{Z} \left(\frac{Z'}{8 \eta \zeta^2  } \mathcal{V}  +\frac{f (n-1) S' \phi '}{\zeta ^2 S} Z'  -\frac{f Z'' \phi'^2}{2 \zeta ^2}\right) +\frac{2 n f  S S' Z' \phi' a'^2}{\mathcal{D} \zeta ^2}  - \\
   &\frac{(n-1) \left(n f' S'-f \eta  S \phi'^2\right)}{\zeta ^2 n S}+\frac{3 f Z'^2 \phi'^2}{4 \zeta ^2 Z^2}\, , \\
W{}^{(0)}_{2,2}(r) &= \frac{k^2}{S^2} + \frac{n-1}{n S^2 \mathcal{D}^2  }\Biggl(
4 n^2 \left(k^2 - n K \right) f S^2 a'^2  Z S'^2  -  8 n \zeta ^4 k^4 K + 8 \eta \zeta^2 f S^2 \phi'^2 k^2 \left(k^2 - n K\right) + \\
&2 n^2 S'^2 \mathcal{F} \left( S f' \left(2 k^2- n K \right)  +  2 f S' \left((n-2)k^2 - n(n-1) K \right)\right)  + \\
&8 \zeta ^2 \left(n S' \left(f S' \left(k^4 - n(n-2) k^2 K + n^2 (n-1) K^2 \right)-k^4 S f'\right) \right)\Biggr) \, ,
\end{split}\end{equation}
where we have further defined,
\begin{equation}\label{eq:DPFAV}
\begin{split}
\mathcal{F}(r) &= 2 f S' - S f' \, , \\
\mathcal{D}(r) &= 2 \zeta ^2 k^2 - n S' \mathcal{F} \, , \\
\mathcal{P}(r) &= \left(\eta  S^2 \phi'^2 - n S'^2\right) \mathcal{F} \, , \\
\mathcal{A}(r) &= 4 n \eta f S'  S^2 Z a'^2 \phi'  \, ,
\end{split}
\end{equation}
and where again primes indicate radial derivatives except when acting on $V$, $Z$ or $\mathcal{V}$, where they indicate a derivative with respect to $\phi$.

The three interaction potentials are,
\begin{equation}\begin{split}\label{eq:potScalar2}
W{}^{(0)}_{0,1}(r) &= - \frac{k \sqrt{Z} a'}{\sqrt{2} \mathcal{D}^2 \zeta  \sqrt{\eta }} \Biggl( 
\mathcal{A} + \mathcal{D} S \mathcal{V}  + \frac{\mathcal{D}^2 Z'}{S Z} + 2\eta \mathcal{D} S   f  \phi'^2 Z'/Z   +  \\
&4 \eta f \phi ' \left(\mathcal{P}  - n(n-1) S'^2 \mathcal{F}- 2 \zeta ^2 S' \left((1-2 n) k^2 + n (n-1) K\right)\right)
\Biggr) \, , \\
W{}^{(0)}_{0,2}(r) &= k \sqrt{k^2- n K} \frac{ \sqrt{n-1} }{ \sqrt{n} \sqrt{\eta } S\mathcal{D}^2} \Biggl(
 \mathcal{A} + \mathcal{D} S \mathcal{V}  + 4 \eta f  \phi ' \left( \mathcal{P} + 2 \zeta ^2 n S' \left(k^2-(n-1)K \right)\right)
\Biggr) \, , \\
W{}^{(0)}_{1,2}(r) &= - \sqrt{k^2- n K} \frac{\sqrt{2} \sqrt{n-1} \sqrt{Z} a' }{ \zeta  \sqrt{n} S \mathcal{D}^2} \Biggl(
2 f n^2 S^2 Z a'^2 S'^2  +  n^2 S f' S'^2 \mathcal{F} +\\
&4 f \zeta ^2 \left(n S'^2 \left(k^2 (n-2)-K (n-1) n\right)+k^2 \eta  S^2 \phi'^2\right)+\mathcal{D} f n S S'  \phi'\frac{Z'}{Z}+4 \zeta ^4 k^4
\Biggr)\, , 
\end{split}\end{equation}

We note again that if either the scalar or the gauge field vanish on the background, their respective master scalars decouple from the rest.
Furthermore at zero momentum in the planar case, all equations decouple.

\subsection{Full decoupling}\label{sec:fulldecoupling}

The final master equations are a single decoupled equation for the tensor sector, two coupled equations for the vector sector and three coupled equations for the scalar sector.

We would like to be able to decouple the vector and scalar sector further to fully decoupled equations. 
If this is possible, the decoupled potentials would be the eigenvalues of the potential matrix,

\begin{equation}
W^{(1)}_\pm = \frac{1}{2} \left( W^{(1)}_{1,1} + W^{(1)}_{2,2} \pm \sqrt{\left(W^{(1)}_{1,1} - W^{(1)}_{2,2}\right)^2 + 4 \left(W^{(1)}_{1,2}\right)^2} \right) \, .
\end{equation}

However, it is only possible to decouple the equations in this way if the eigenvectors of the potential matrix do not depend on $r$.
Computing the eigenvalues and taking the $r$-derivative, one finds that the equations can be decoupled under the condition that:
\begin{equation}
\partial_r \log \left( W^{(1)}_{1,2} \right)= \partial_r \log \left(W^{(1)}_{1,1} - W^{(1)}_{2,2} \right)
\end{equation}

More simply, they can be decoupled if
\begin{equation}\label{eq:constancycriterion}
\frac{W^{(1)}_{1,1} - W^{(1)}_{2,2}}{W^{(1)}_{1,2}} = \text{const.} \,  ,
\end{equation}
and in that case, the $r$-dependence in the square root in the eigenvalues factors out, leaving a square root only of constants.

In the sound channel the algebra is a bit more complicated, with the decoupled potentials being the following eigenvalues of the potential matrix (\textit{if} it can be decoupled):
\begin{equation}
W^{(0)}_\sigma = \frac{1}{3} \left( \mathcal{T} + \sigma \frac{\mathcal{T}^2 + 3 \mathcal{U} }{\mathcal{C}} + \frac{\mathcal{C}}{\sigma} \right) \, ,
\end{equation}
where $\sigma$ are the three roots of $\sigma^3 = 2$, and
\begin{equation}\begin{split}
\mathcal{T} &= \mathfrak{tr} \left( W^{(0)} \right) \, , \\
\mathcal{U} &= \frac{1}{2}\left(\mathfrak{tr} \left( \left(W^{(0)}\right)^2 \right) -  \mathcal{T}^2 \right) \, , \\
\mathcal{D} &= \text{det} \left( W^{(0)} \right) \, , \\
\mathcal{C} &=  \left( 27 \mathcal{D} + 2 \mathcal{T}^3 + 9 \mathcal{T} \mathcal{U} + \mathcal{R} \right)^{1/3} \, , \\
\mathcal{R} &=  \left( \left( 27 \mathcal{D} + 2 \mathcal{T}^3 + 9 \mathcal{T} \mathcal{U}  \right)^2 - 4 \left( \mathcal{T}^2 + 3 \mathcal{U}^2\right)^3  \right)^{1/2} \, . \\
\end{split}\end{equation}

Again this decoupling can only be done when the eigenvectors are constant, we have however not been able to derive a simple criterion such as Eq. (\ref{eq:constancycriterion}) in this case.

\subsection{Comparison to Kodama-Ishibashi}
To compare with the results of Kodama and Ishibashi~\cite{Kodama:2003kk} for Reissner-Nordstr\"{o}m we turn off the scalar field,
\begin{equation}
\phi(r) = 0 \, , \quad V(\phi) = 0 \, , \quad Z(\phi) = 1 \, ,
\end{equation}
and we insert the Reissner-Nordstr\"{o}m solution,
\begin{equation}\begin{split}
\zeta(r) &= 1 \, , \\
S(r) &= r \, , \\
a(r) &= \sqrt{\frac{2n}{n-1}} Q r^{1-n} \, , \\
f(r) &= K - \lambda r^2 - \frac{2 M}{r^{n-1}} + \frac{Q^2}{r^{2n-2}} \, . \\
\end{split}\end{equation}

From the decoupling condition Eq. (\ref{eq:constancycriterion}) we see by inserting this background that the equations can be decoupled:
\begin{equation}
\frac{W^{(1)}_{1,1} - W^{(1)}_{2,2}}{W^{(1)}_{1,2}}  = \frac{1}{\sqrt{k^2-K n}} \frac{\left(n^2-1\right) }{ \sqrt{2 n (n-1)} } \frac{\left(K-\lambda +Q^2\right)}{Q} \, ,
\end{equation}
and since we've set the scalar to zero the scalar sector also has only two equations, so we can apply the same criterion and find:
\begin{equation}
\frac{W^{(0)}_{1,1} - W^{(0)}_{2,2}}{W^{(0)}_{1,2}}  = \frac{1}{\sqrt{k^2-K n}} \frac{n+1}{ 2 } \frac{\left(K -\lambda +Q^2\right)}{Q} \, .
\end{equation}

So the equations can indeed be fully decoupled and the resulting potentials are the eigenvalues of our potential matrices.

In order to compare these with KI we first have to transform them to the Schr\"{o}dinger form,
\begin{equation}\label{eq:schrodingerPotential}
V_S(r) = W(r) + \left(\frac{n}{4} \frac{S'}{S^2 \zeta^2} \left( 2 S f' + (n-2) f S' \right)- \frac{n}{2} \frac{f \zeta'}{\zeta^3} \frac{S'}{S} + \frac{n}{2} \frac{f}{\zeta^2} \frac{S''}{S} \right)  \mathbb{1}
\, .
\end{equation}
We will see how this arises in the next section, and note that this redefinition drops out in the condition of Eq. (\ref{eq:constancycriterion}).

Computing the eigenvalues, which we shall call $\tilde{W}^{(h)}$, from these potential matrices in Schr\"{o}dinger form, we obtain for the tensor sector:
\begin{equation}
\tilde{W}^{(2)} = \frac{1}{4 r^{2(n+1)}} \left(r^{2 n} \left(4 k^2+K (n-2) n-\lambda  n (n+2) r^2\right)+2
   M n^2 r^{n+1}+(2-3 n) n Q^2 r^2\right) \, ,
\end{equation}
which agrees with Eq. (3.7) in~\cite{Kodama:2003kk} (with $\lambda_L = k^2 + 2(n-1) K$).

In the vector sector we obtain the two eigenvalues:
\begin{equation}\begin{split}
\tilde{W}^{(1)}_\pm &= \frac{1}{4} r^{-2 (n+1)} \left(r^{2 n} \left(4 k^2+(n-2) n \left(K-\lambda 
   r^2\right)\right)-2 M \left(n^2+2\right) r^{n+1}+n (5 n-2) Q^2 r^2\right) \\
   & \pm \frac{\Delta^{(1)}}{ r^{(n+1)}} \, , \\
\Delta^{(1)} &= \sqrt{(n^2-1)^2 M^2  + 2 n (n-1) (k^2 - n K) Q^2} \, ,
\end{split}\end{equation}
which agrees with Eq. (4.38) in~\cite{Kodama:2003kk} (with $k_V = k^2 - K$).

Finally in the scalar sector we obtain two significantly more complicated eigenvalues:
\begin{equation}\label{eq:KIscalar}\begin{split}
\tilde{W}^{(0)}_\pm &= \tilde{W}^{(0)}_1(r) \pm \Delta^{(0)} \tilde{W}^{(0)}_2(r)\, , \\
\Delta^{(0)} &= \sqrt{(n^2-1)^2 M^2  + 4 (n-1)^2 (k^2 - n K) Q^2}\, ,
\end{split}\end{equation}
with $\tilde{W}^{(0)}_{1,2}$ functions too long to reproduce here.

This again agrees with KI, Eq. (5.61 - 5.63), although superficially they appear very different, in particular the structure of Eq. (\ref{eq:KIscalar}) is not visible in~\cite{Kodama:2003kk}.

\section{Stability}\label{sec:stability}

If we define $\Phi_s^{(h)}(t,r)= e^{- i \omega t } S(r)^{- n /2} \Psi_s^{(h)}(r)$, and evaluate Eq. (\ref{eq:masterequation}) in Eddington-Finkelstein coordinates (\ref{eq:EFansatz}), we obtain the following Schr\"{o}dinger-like equation,
\begin{equation}\label{eq:schrodinger}
X \equiv \partial_r \left( \frac{f}{\zeta} \partial_r \Psi(r) \right) - 2 i \omega \partial_r \Psi(r) -  \zeta V_S(r) \Psi(r) = 0 \, ,
\end{equation}
where for simplicity we drop $s$ indices and $h$ labels, but $\Psi$ is still a vector with 1, 2 or 3 components for the tensor, vector and scalar channel respectively, 
and $V_S$ is the corresponding Schr\"{o}dinger potential matrix, that is related to the original potential $W$ in Eq.~(\ref{eq:masterequation}) as in Eq.~(\ref{eq:schrodingerPotential}).

From the Schr\"{o}dinger-like equation (\ref{eq:schrodinger}) it is possible to derive a sufficient, but not necessary, condition for linear stability of the corresponding fluctuation~\cite{Horowitz:1999jd}.
We review the argument here.

Start by defining the vanishing integral,
\begin{equation}
I \equiv - \int_{r_h}^{\infty} \bar{\Psi} X  =  \int_{r_h}^{\infty} \left( - \bar{\Psi} \partial_r \left( \frac{f}{\zeta} \partial_r \Psi \right) + 2 i \omega \bar{\Psi} \partial_r \Psi + \zeta \bar{\Psi} V_S \Psi \right)\, .
\end{equation}

By partial integration, this can be written as
\begin{equation}
I  = \int_{r_h}^{\infty} \left(  \frac{f}{\zeta} |\partial_r \Psi|^2+ 2 i \omega \bar{\Psi} \partial_r \Psi + \zeta \bar{\Psi}  V_S  \Psi \right) - \frac{f}{\zeta} \bar{\Psi} \partial_r\Psi |_{r_h}^{\infty}  \, .
\end{equation}

Provided $\Psi$ is regular at the horizon and dies off sufficiently fast at infinity, which are exactly the conditions for quasinormal modes, the boundary term vanishes.

From the above we obtain,
\begin{equation}
\text{Im}(I) = 0 = \int_{r_h}^{\infty} \left(\omega \bar{\Psi} \partial_r \Psi + \bar{\omega} \Psi \partial_r \bar{\Psi} \right) \, , 
\end{equation}
where we have used that $W$, and thus $V$, is a real and symmetric matrix.

Now integrating the last term by parts we get
\begin{equation}
\left( \omega - \bar{\omega} \right) \int_{r_h}^{\infty}  \bar{\Psi} \partial_r \Psi = \bar{\omega} |\Psi(r_h)|^2 - \bar{\omega} | \Psi(\infty)|^2 \, ,
\end{equation}
where the last term vanishes again assuming that $\Psi$ dies off sufficiently fast.

Inserting this into $I$ we finally obtain:
\begin{equation}
J \equiv \int_{r_h}^{\infty} \left(  \frac{f}{\zeta} |\partial_r \Psi|^2 + \zeta \bar{\Psi} V_S  \Psi \right) = - \frac{|\omega|^2}{\omega_I} |\Psi(r_h)|^2 \, .
\end{equation}

From this we see that $\omega_I$ is negative, meaning the perturbation is stable, if and only if the integral $J$ is positive. 
Since we do not know $\Psi(r)$, this is not directly useful.
However a sufficient condition is that the eigenvalues of $V$ are positive everywhere outside of the black hole.

We stress that it is not required that although in practice it helps if the equations can be further decoupled, it is not necessary for this argument.
The only requirement is that the potential matrix be symmetric (or more generally Hermitian), which it explicitly is for any theory within our setup.

\subsection{$\mathcal{S}$-deformation}
We can get something more by transforming the integral with what is called an $\mathcal{S}$-deformation~\cite{Kodama:2003kk}.

For some arbitrary, possibly matrix valued, function $\mathcal{S}$, define
\begin{equation}\begin{split}
\tilde{D} &\equiv \partial_r + \frac{\zeta}{f} \mathcal{S} \, , \\
\tilde{V}_S &\equiv V_S + \frac{1}{\zeta} \left( \mathcal{S}^\prime - \frac{\zeta}{f} \mathcal{S}^2 \right) \, , \\
\tilde{J} &\equiv \int_{r_h}^\infty dr \left(\frac{f}{\zeta} |\tilde{D} \Psi |^2 + \zeta \bar{\Psi} \tilde{V}_S \Psi \right) \, .
\end{split}\end{equation}

Provided the boundary term $S |\Psi|^2 |_{r_h}^\infty$ vanishes and $\mathcal{S}$ is real and symmetric, or more generally a Hermitian matrix, $\tilde{J} = J$.

So if we can find any $\mathcal{S}$-deformation that makes the deformed potential positive everywhere, the corresponding perturbation is stable.

In practice, if the system can be decoupled it is usually easier to first decouple and then find an $\mathcal{S}$-deformation for the decoupled potentials.
However, it is also possible, and indeed if they do not decouple the only way, to deform the potential matrix with a Hermitian matrix $\mathcal{S}$, and then try to show positivity of the eigenvalues of the deformed potential matrix.

This was used in~\cite{Ishibashi:2003ap} to prove stability of Reissner-Nordstr\"{o}m black holes.
If an analytic $\mathcal{S}$-deformation cannot be found one can also look for a numerical $\mathcal{S}$-deformation that is regular and makes the transformed potential vanish \cite{Kimura:2018whv}.

In the following sections we find analytic analytic $\mathcal{S}$-deformations to prove stability for various specific cases.
\subsection{Stability of tensor perturbations}

The tensor perturbations are the simplest of all, having a potential $W^{(2)}$ that comes only from the eigenvalue of the Laplacian, with the additional contribution of Eq. (\ref{eq:schrodingerPotential}).

We can deform this with the $\mathcal{S}$-deformation
\begin{equation}
\mathcal{S}^{(2)} = - \frac{n}{2} \frac{f S'}{\zeta S} \, ,
\end{equation}
to obtain the manifestly positive deformed potential:
\begin{equation}
\tilde{V}^{(2)}_S = \frac{k^2}{S^2} \, .
\end{equation}

Hence the tensor modes are always stable.

\subsection{Stability of vector perturbations in Einstein-scalar theory}

The vector sector is already significantly more complicated and we cannot prove stability in general.

However for specific cases we can, and curiously in the cases where it can be done, it can be done with the same simple $\mathcal{S}$-deformation:
\begin{equation}\label{eq:SdeformH1}
\mathcal{S}^{(1)} = + \frac{n}{2} \frac{f S'}{\zeta S} \, ,
\end{equation}
the negative of the tensor one.

In particular if we turn off the gauge field, leaving just Einstein-scalar theory with an arbitrary potential, we are left with a single decoupled equation and the rather simple potential $W^{(1)}_{2,2}$.
The modified potential becomes,
\begin{equation}
\tilde{V}^{(1)}_S = \frac{k^2 - n K}{S^2} \, .
\end{equation}

This is manifestly positive for $K = 0$ and $-1$, and in the spherical case $k^2 = l(l+n - 1)$, so the numerator becomes $(l-1)(l+n)$, also manifestly positive since in the vector sector $l \geq 1$.

\subsection{Stability of vector perturbations of the GMGHS black hole}\label{sec:strominger}
We now turn to a more involved application that has all the fields in our ansatz, an asymptotically flat charged black hole with a dilaton in 3+1 dimensions, which minimizes the action given by our ansatz (\ref{eq:action}), with
\begin{equation}\begin{split}
\eta &= 2  \, , \\
V(\phi) &= 0 \, , \\
\Lambda &= 0 \, , \\
Z(\phi) &= 4 e^{-2 \alpha \phi} \, ,
\end{split}\end{equation}
where $\alpha$ is a free parameter corresponding to the dilaton coupling.
For $\alpha = 0$ this reduces to the usual Reissner-Nordstr\"{o}m action, while for $\alpha = 1$ this is the low energy effective action obtained from heterotic string theory.

The solution is given by \cite{Gibbons:1987ps,PhysRevD.43.3140},
\begin{equation}\begin{split}
f(r) &= \left(1 - \frac{R_+}{r}\right) \left(1 - \frac{R_-}{r}\right)^{\frac{1 - \alpha^2}{1 + \alpha^2}}\, , \\
\zeta(r) &= 1 \, , \\
S(r) &= r \left(1 - \frac{R_-}{r}\right)^{\frac{\alpha^2}{1 + \alpha^2}} \, , \\
a(r) &= \sqrt{\frac{R_+ R_-}{1 + \alpha^2}} \frac{1}{r}  \, , \\
e^{\alpha \phi(r)} &= \left(1 - \frac{R_-}{r}\right)^{\frac{\alpha^2}{1 + \alpha^2}} \, , 
\end{split}\end{equation}
where $R_+ \geq R_-$ are the horizon and singularity respectively, in which the charge and mass can be expressed as: $2 M = R_+ + \frac{1 - \alpha^2}{1 + \alpha^2 } R_-$ and $Q^2 = \frac{R_+ R_-}{1 + \alpha^2}$.

For this solution, the master equations can be completely decoupled. Indeed this was already noted in \cite{Holzhey:1991bx}, where stability was also argued for by numerical inspection of the decoupled potentials for several parameter values, although without analytical proof. 
Perturbations of this background were also analysed in the small charge approximation in~\cite{Brito:2018hjh}.

Here we can prove stability analytically in the vector sector with the same $\mathcal{S}$-deformation of Eq. (\ref{eq:SdeformH1}), and the process is only slightly more involved.
The deformed potential takes the form,
\begin{equation}\begin{split}
\tilde{V}^{(1)}_{S \pm} &= \frac{\left(1 - R_- / r \right)^{\frac{2}{1+ \alpha^2}}}{2  r (r - R_-)^2} \biggl( 2 R_- \left(1 + \delta l \left(3 + \delta l \right) \right) + 3 \delta R + 2 \delta l \left(3 + \delta l \right) \left( \delta r + \delta R \right) \pm \Delta  \\
&+ \frac{4 R_-}{1+ \alpha^2} \biggr) \, ,
\end{split}\end{equation}
where
\begin{equation}
\Delta^2 = \frac{16 \delta l (\delta l+3) \left(\alpha ^2+1\right) R_- (\delta R+R_-)+\left(3 \delta R \left(\alpha ^2+1\right)+2 \left(\alpha ^2+3\right)
   R_-\right)^2}{\left(\alpha ^2+1\right)^2} \, ,
\end{equation}
and to be able to more easily show positivity we have defined the manifestly positive, or at least non-negative, quantities:
\begin{equation}\begin{split}
\delta R &= R_+ - R_- \, , \\
\delta r &= r - R_+ \, , \\
\delta l &= l - 1 \, .
\end{split}\end{equation}

Now in $\tilde{V}^{(1)}_{S +}$ every symbol is positive, and there are no minus signs in the expression, so this is manifestly positive.

To show the same for $\tilde{V}^{(1)}_{S -}$, we have to show that $\Delta$ is smaller than the sum of the other terms in the expression.
Or equivalently, we can show that $\Delta^2$ is smaller than the square of the sum of the remaining terms. 
Simply writing this out using the same definitions as above, this is immediately seen to be true.

In the scalar channel the equations can also be decoupled. However the resulting potentials are very complicated and we have not been able to do a similar analytic stability proof in this case.

\section{Discussion}

We have reduced the problem of linear fluctuations in Einstein-Maxwell-scalar theories with maximally symmetric horizons to a small set of master scalars, one for each graviton polarization, in which everything can be expressed analytically.
The equations fall into three sectors, tensor, vector and scalar, consisting of respectively a single decoupled equations and 2 and 3 coupled equations.

Although the potentials in the resulting master equations are rather complicated, the form is conceptually very simple.
In fact it is not clear to us why we could obtain such a simple form, in particular with symmetric potential matrices in the coupled equations.

Furthermore, in several cases, such as Reissner-Nordstr\"{o}m and the GMGHS black hole of \cite{Gibbons:1987ps,PhysRevD.43.3140}, the coupled equations can be decoupled further into fully decoupled equations.
However there also exist analytic solutions, such as the asymptotically Lifshitz black brane of \cite{Tarrio:2011de}, where this cannot be done.
It is not clear to us on a general level what distinguishes these theories.

The symmetry of the potential matrices allows one to derive a sufficient condition for stability, and the full decoupling makes it simpler to apply.

Furthermore our way of deriving the master equations is conceptually very simple and we believe would be rather simple to generalize to for instance other matter content. One simply has to find an ansatz which is sufficiently, but not too, general and ask Mathematica nicely to solve it for you.

We expect that our results can be straightforwardly generalized to include also time-dependence in the background, as was done in~\cite{Kodama:2003kk,Rostworowski:2017ruj,Rostworowski:2019iig}

Finally we wish to comment on the differences with the Kovtun-Starinets (KS) approach \cite{Kovtun:2005ev} to solve fluctuation equations for black-branes, a widely used approach in holography. In appendix \ref{app:KS} we go into more detail.
Instead of expressing all gauge-invariants in terms of a master scalar, KS single out one gauge-invariant, $\mathfrak{h}_{tt}$, and decouple its equation from the others.
We believe that the Kodama-Ishibashi approach that we follow has several advantages. 
On a more conceptual level, one explicitly solves all the equations of motion by solving the master scalar equations, 
and one can reconstruct analytically all of the components of the metric and the matter fields. Furthermore the master equations one obtains are covariant.
On a practical, numerical, level, we find that the master equations are more accurate in finding the quasinormal modes in several ways.
We discuss this in appendix \ref{app:KS}, along with a quantitative comparison.

In holographic studies a full quasinormal mode analysis is often lacking, especially in the most complicated sound channel. We hope that this work simplifies this sufficiently to make such a complete analysis more accessible and thus more common.

\section{Acknowledgements}
The authors are indebted to Jorge Rocha for careful reading of the early version of the manuscript and many useful comments. AJ and AR  wish to thank Pavel Kovtun for a helpful discussion, and AJ wants to thank Vitor Cardoso and Caio Macedo for helpful discussions.
The work of AJ is supported by ERC Advanced Grant GravBHs-692951.
The work of AR and MR is supported by the Narodowe Centrum Nauki (Republic of Poland) Grant no. 2017/26/A/ST2/530.
AJ and AR are grateful to Juan Pedraza for organizing the \textit{Aspects of Time Dependent Holography} workshop held in Amsterdam, 4-8.12.2017, where the project resulting in the present work began.

\bibliographystyle{jhep}                 
\bibliography{references}

\appendix

\section{Spherical Case}\label{app:spherical}
A convenient way to parametrize all three $n$-dimensional maximally symmetric spaces at once is as $X = (x_{1} \equiv x,x_{2} \equiv  y,x_{3}, ..., x_{n-1}, x_{n} \equiv  z)$ and
\begin{equation}\begin{split}
dX_{(1)}^2 &= d x_n^2 \, , \\
dX_{(i)}^2 &= \frac{1}{\left(1 - K x_{n-i+1}^2\right)} d x_{n-i+1}^2 + \left(1 - K x_{n-i+1}^2\right) dX_{(i-1)}^2 \, ,
\end{split}\end{equation}
where $i = 2, ..., n$ and 
\begin{equation}
K =  
\begin{cases}
+1 & \text{spherical} \\
\,\,\,0 &\text{planar} \\
-1& \text{hyperbolic} 
\end{cases} \, .
\end{equation}

The difficult part of the spherical case with respect to the planar is the decomposition into the three sectors.
That is, to find an ansatz for the fluctuations analogous to Eq. (\ref{eq:components}),  in such a way that the decomposition given in Table \ref{tab:sectors} still applies.
Note that the very simple ansatz in the planar case no longer suffices because under a rotation components get mixed.

Once such an ansatz has been found, everything else goes through in exactly the same manner, so here we discuss this ansatz.

To find this ansatz we follow the treatment of the harmonics on maximally symmetric spaces in Appendix B of \cite{Mukohyama:2000ui}.

The basic ingredient for constructing the different components is of course the scalar eigenfunction $S$ of the Laplacian on the maximally symmetric space $D^2$, with eigenvalue $k^2$,
\begin{equation}
D^2 S + k^2 S = 0 \, .
\end{equation}
We can choose $S$ that depends only on $x$, and the equation becomes
\begin{equation}
\left( 1 - K x^2 \right) S''(x) = n K x S'(x) - k^2 S(x) \, ,
\end{equation}
with $k^2 = l (l + n - 1)$.

With this we can immediately write the $(t,r)$ part of the fluctuations as ($a,b \in \{t,r\}$)
\begin{equation}\label{eq:metricSphericaltr}
\delta g_{a b} = 
\begin{pmatrix}
h_{t t} 			&1/2 h_{t r} 		\\
1/2 h_{t r} 			& h_{r r}			\\
\end{pmatrix}  S(x) \, , 
\end{equation}
this piece can remain unchanged apart from the change of the scalar eigenfunction $S(x)$.

Any vector can be decomposed into a longitudinal and transverse part,
\begin{equation}\begin{split}
V &= V_L + V_T  \, , \\
D^{i} V_{T,i} &= 0 \, , \\
V_{L,i} &= \partial_i S \, .
\end{split}\end{equation}

For the longitudinal part we already have an explicit expression in terms of $S$, and furthermore since $S$ depends only on $x$ it reduces to a single component. 
This will also contribute to the scalar sector and can be used to express $h_{t x}$ and $h_{r x}$.

The transverse part will contribute to the vector sector and is not readily expressible in terms of the scalar $S$.
It must however satisfy the equation (as does the longitudinal part),
\begin{equation}
D^2 V_T + \left(k^2 - K\right) V_T = 0 \, ,
\end{equation}
with a shifted eigenvalue.

As in the planar case there are $n-1$ solutions, but it suffices to find a single one, since by symmetry all should satisfy the same equations.
The simplest solution to this equation is,
\begin{equation}\begin{split}
V_{T,n} &= S_V(x) \, , \quad V_{T,i} = 0 (i \neq n) \, , \\
\left( 1 - K x^2 \right) S_V''(x) &= (n - 2) K x S_V'(x) - \left(k^2 + (n-2) K \right) S_V(x) \, .
\end{split}\end{equation}
Note that we had to choose the last component to be nonzero in order to avoid explicit dependence on the other coordinates on the sphere.

With these we can express the $a i$ part of the metric fluctuations, where $a \in \{t,r\}$ and $i$ goes over the remaining coordinates, as,
\begin{equation}\label{eq:metricSphericalti}
\delta g_{a i} = 
\begin{pmatrix}
h_{t x} S'(x) 	& 0 & ... & 0 & h_{t z} S_V(x)	\\
h_{r x} S'(x)	& 0 & ... & 0 & h_{r z} S_V(x)	\\
\end{pmatrix} \, , 
\end{equation}

and any symmetric tensor can be decomposed into a transverse, traceless and symmetric part, a longitudinal part and a trace part,
\begin{equation}\begin{split}
T &= T_{TT} + T_L + T_T  \, , \\
T_{T,ij} &= S \Omega_{ij} \, , \\
T_{L,ij} &= D_i V_j + D_j V_i \, , \\
D^i T_{TT,ij} &= 0 \, , \\
T_{TT,i}^i &= 0 \, .
\end{split}\end{equation}
So we already have explicit expressions for the trace and longitudinal part. The former contributes to the scalar sector through $h_+$, and the latter comes in two parts since for the vector it's built on we can take the longitudinal or transverse vector.
It is more convenient to redefine these as:
\begin{equation}\begin{split}
T_{LT,ij} &= D_i V_{T,j} + D_j V_{T,i} \, , \\
T_{LL,ij} &= D_i V_{L,j} + D_j V_{L,i} - \frac{2}{n} D^{k} V_{L,k}  \Omega_{ij}\, . \\
\end{split}\end{equation}
Here $T_{LT}$ will contribute to the vector sector through $h_{xz}$, and $T_{LL}$ will contribute to the scalar sector through $h_-$.

The transverse traceless part must satisfy the equation (as do the other components),
\begin{equation}
D^2 T_{TT} + \left(k^2 - 2 K\right) T_{TT} = 0 \, ,
\end{equation}

Again it suffices to find a single solution to this equation, which we have found to be,
\begin{equation}\begin{split}
T_{TT,yz} = T_{TT,zy}  &= \left(1 - K y^2 \right)^{-(n-1)/2} \left( \prod\limits_{i=3}^{n-1} (1 - K x_i^2 ) \right) S_T(x) \, , \quad (T_{T,ij} = 0 \,\, \text{otherwise})) \, , \\
\left( 1 - K x^2 \right) S_T''(x) &= (n - 4) K x S_T'(x) - \left(k^2 +2 K \left( 1 - (n-3) \frac{K x^2}{1 - K x^2}\right) \right) S_T(x) \, .
\end{split}\end{equation}
This component is in the tensor sector.

Summarizing, the full metric perturbation we do is,
\begin{equation}\begin{split}
\delta g_{\mu\nu} &= \left(h_{t t} dt^2 + h_{t r} dt dr + h_{r r} dr^2 + \frac{1}{n} h_+ \Omega_{ij} dx^i dx^j  \right) S + 2 \left( h_{t x} dt dx +  h_{r x}  dr dx \right) S'(x) \\
&+ 2 \left(h_{t z}  dt dz  + h_{r z}  dr dz  \right) S_V + \left(\frac{1}{2} h_- T_{LL,ij} + h_{x z} T_{LT,ij}  + h_{y z} T_{TT,i j} \right)dx^i dx^j  \, ,
\end{split}\end{equation}
and this reduces exactly to Eq. (\ref{eq:components}) when $K = 0$.

\section{Special cases}

In this appendix we discuss the special cases, which are $l = 1$ and $l = 0$ in the spherical case and $k = 0$ in the planar case.
Although these cases are special, at the end of the day if one is interested in the quasinormal mode spectrum, one can use the potentials derived for the general case here as well.

\subsection{Spherical $l = 1$ in vector sector}\label{app:specialCasesVector}
In the following derivation, we follow \citep{Bicak:1979}. For $l = 1$, there is no $h_{xz}$ in the harmonic decomposition, therefore $\mathfrak{h}_{t z}$ and $\mathfrak{h}_{r z}$ defined in \eqref{eq:gaugeInvariants} are not gauge invariant anymore. We can fix the gauge to set $h_{r z} = 0$. Then, from equations $E_{tz} = 0$ and $E_{rz} = 0$, we find:
\begin{equation}
\begin{split} 
\partial_r\left(\frac{h_{t z}}{S^2}\right)&= - \frac{\sqrt{Z} a' }{S}  \Phi^{(1)}_1 - c_1 \frac{ \zeta}{S^{n+2}}\, ,\\
a_{z}& = \frac{S }{\sqrt{Z}}\Phi^{(1)}_1  \, ,
\end{split}
\end{equation}
$c_1$ being an arbitrary constant. $\Phi^{(1)}_1$ fulfils an \textit{inhomogeneous} wave equation:
\begin{equation}\label{eq:inhomog}
\square \Phi^{(1)}_1 - W{}^{(1)}_{1,1} \Phi^{(1)}_{1} = c_1 \frac{  \sqrt{Z} a'}{\zeta  S^{n+1}} \, ,
\end{equation}
where potential $W{}^{(1)}_{1,1}$ is given by \eqref{eq:potScalar1} with $K=1$ and $k = \sqrt{n}$ ($l =1 $). The interpretation of \eqref{eq:inhomog} is the following: particular (stationary) solutions contribute to the angular momentum of the black hole (e.g. to linearised Kerr--Newman black hole in 3+1 dimensions), whereas the homogeneous solution is the dynamical degree of freedom of an electromagnetic field.

\subsection{Spherical $l = 0$ and $l = 1$ in scalar sector}\label{app:specialCasesScalar}
At $l = 0$ (with $K = 1$, so $k = 0$) the only dynamical degree of freedom is in the scalar field. Naively plugging this in into our potentials, one sees that now the interaction terms between the scalar master scalar and the others, $W^{(0)}_{0,1}$ and $W^{(0)}_{0,2}$ vanish. So one is left with a decoupled master equation for the only physical degree of freedom which is in the scalar field, and the potential is simply the one we found, $W^{(0)}_{0,0}$, which is perfectly regular for $k = 0$. 

To obtain this result, we firstly use the fact that for $l=0$ there are no $h_{tx}$, $h_{rx}$, $h_-$ and $a_x$ components of perturbations and no $\xi_x$ gauge vector component. Gauge invariants defined for $k^2>n$ do not make sense anymore. Instead, we can use $\xi_t$, $\xi_r$ and $\lambda$ to set e.g.\ $h_{tr}$, $h_+$ and $a_t$ to zero. Making such a choice, we are left with four variables: $h_{tt}$, $h_{tr}$, $a_r$, $\varphi$. As expected, scalar field is the only dynamical variable in the system:  
\begin{equation}
\varphi = \Phi^{(0)}_0\, .\\
\end{equation}
This master scalar satisfies the wave equation with the potential $W^{(0)}_{0,0}$ of the general case, with $l = 0$ plugged in, but, as well as in the vector $l=1$ case, this wave equation is inhomogeneous:
\begin{equation}\label{eq:inhomog2}
\square \Phi^{(0)}_0 - W{}^{(0)}_{0,0} \Phi^{(0)}_{0} = \frac{c_0 (a'^2 Z'-2 \zeta ^2 V')}{4 f \zeta  \eta  S^{n-1} S'}+\frac{c_0 \left(S f'+f (n-1) S'\right)\phi ' }{f \zeta  S^n S'}-\frac{c_0 \eta  \phi '^3}{\zeta  n S^{n-2} S'^2}\, ,
\end{equation}
$c_0$ being an arbitrary constant. The other fluctuations can be found from the Einstein equations directly. In contrary to $l\geq2$, however, not all of them can be directly expressed by a master scalar and it's derivatives, but integration for $h_{tt}$ and $a_r$ will be necessary:
\begin{equation}\begin{split} 
h_{rr} &=\frac{c_0 \zeta^3}{f^2 S^{n-1} S'}+ \frac{2 \zeta ^2 \eta  S \phi '}{f n S'} \Phi^{(0)}_0\, ,\\
f \partial_r\left(\frac{h_{tt}}{f}\right) &=\left(\frac{S \left(2 \zeta ^2 V'- Z' a'^2-4 \eta  f' \phi '\right)}{2 n S'}+\frac{2 f \eta ^2 S^2 \phi '^3}{n^2 S'^2}-\frac{2 (n-1) \eta  f\phi'}{n}\right)\Phi^{(0)}_0+\\
-&\frac{2 f \eta  S \phi ' }{n S'}\partial_r \Phi^{(0)}_0+\frac{c_0 \zeta  \left(f \eta  S^2 \phi '^2-n S f' S'-f n(n-1) S'^2\right)}{f n S^n S'^2},\\
\partial_t a_r &=\frac{a' h_{tt}}{2 f}+\frac{1}{n} a'  \left(\frac{n Z'}{Z}-\frac{\eta  S \phi '}{S'}\right)\Phi^{(0)}_0-\frac{c_0 \zeta  a'}{2 f S^{n-1} S'}\, .
\end{split}\end{equation}
The constant $c_0$ corresponds to a static perturbation of the zeroth order solution (e.g.\ to a shift of mass in the Reissner-Nordstr\"{o}m background).

The case $l = 1$ (with $K = 1$, so $k^2 = n$) is special because since the metric has spin 2, it does not have any dynamical degrees of freedom with $l = 1$.
This is the reason that we see factors of $\sqrt{k^2 - n K}$ in the potentials. 
In particular this factor occurs as a prefactor in all interaction potentials of the gravitational master scalar with the others, so $W^{(1)}_{1,2}$, $W^{(0)}_{0,2}$ and $W^{(0)}_{1,2}$.
So simply plugging in $l = 1$ in the potentials we have the master equations for the scalar and gauge field fluctuations, which do have physical degrees of freedom with $l = 1$, and they decouple from the unphysical gravitational degrees of freedom.

To see this more concretely, note that for $l = 1$ there is no $h_-$ component (the spherical harmonic can be explicitly solved in this case to be $S(x) = x$, and the $T_{LL}$ tensor that defines $h_-$ is given by second derivatives of this, hence vanishing). 
This means that the gauge-invariants of \eqref{eq:gaugeInvariants} are no longer gauge invariant. In particular they transform under the infinitesimal coordinate transformation component $\xi_x$ as, 
\begin{equation}\begin{split} 
\mathfrak{h}_{t t} &\rightarrow \mathfrak{h}_{t t}  +\frac{f'   }{S  S' }\xi _x+2 \partial_t^2\xi _x\, , \\
\mathfrak{h}_{t r} &\rightarrow \mathfrak{h}_{t r} +2 \partial_t\partial_r\xi _x-\frac{2  \left(S  f'  S' +\zeta^2\right)}{f  S  S' }\partial_t\xi _x\,  , \\
\mathfrak{h}_{r r} &\rightarrow\mathfrak{h}_{r r} +\frac{\zeta ^2 \left(S f' S'+2 f S'^2-\frac{2 f \eta  S^2 \phi '^2}{n}\right)}{f^2 S^2 S'^2} \xi _x  -\frac{2 \zeta^2 }{f  S  S' }\partial_r\xi _x \,, \\
\mathfrak{h}_{r x} &\rightarrow \mathfrak{h}_{rx} + \frac{  \left(2 {S'}^2-\frac{\zeta^2}{f }\right)}{S  S' }\xi _x-\partial_r\xi _x \, , \\
\mathfrak{a}_{t} &\rightarrow \mathfrak{a}_{t}   -\frac{a'   }{S  S' }\xi _x  \, , \\
\mathfrak{a}_{r} &\rightarrow \mathfrak{a}_{r}   +\frac{a' }{f }   \partial_t\xi _x \, , \\
\text{\boldmath$\varphi$}&\rightarrow \text{\boldmath$\varphi$} -\frac{\phi '}{S  S' }   \xi _x\, ,
\end{split}\end{equation}
but are still invariant under the other components.

So we have an extra gauge choice that we are free to make. If we take the simple $\xi _x = -S^2 \Phi^{(0)}_2$ and plug this into equations (\ref{eq:scalarMasterScalars1}) to (\ref{eq:scalarMasterScalars4}) expressing the gauge-invariants in terms of the master scalars, 
we see that $\Phi^{(0)}_2$ completely drops out from all expressions. This confirms the fact that there are no dynamical degrees of freedom in the metric with $l = 1$.

\subsection{Planar $k = 0$}
Without momentum there is no way to distinguish between what at finite momentum were different channels, and so by symmetry one expects the physics to be the same in all channels.
This is seen explicitly in the approach of Kovtun and Starinets~\cite{Kovtun:2005ev}, where the equations for metric fluctuations become identical in each channel, and those for vector perturbations become identical to each other as well.


In our results, setting the momentum to zero makes all the equations decouple.
However, the potentials for a given field are not all identical. In particular, the potential for the metric master scalar in the scalar channel  vanishes, but in the vector channel it does not. The gauge field potentials do not vanish and are not equal to each other either.

This is not what we expect by symmetry, but there is an elegant solution.
Defining Schr\"{o}dinger-like potentials in Fefferman-Graham or Schwarzschild-like coordinates as\footnote{This makes $\tilde{V}_S = f(r) V_S$ with $V_S$ defined in Eq. (\ref{eq:schrodingerPotential}).},
\begin{equation}
\frac{d^2 \Psi}{d r_\star^2} + \left( \omega^2 - \tilde{V}_S \right) \Psi = 0 \, ,
\end{equation}
with $\partial_{r_\star} = \frac{f}{\zeta} \partial_r$, then if two of these potentials can be written in terms of a single super potential $W_S$ as
\begin{equation}
\tilde{V}_{S,\pm} = W_S^2 \mp \frac{d W_S}{d r_\star} + \beta \, ,
\end{equation}
then these two potentials are isospectral, having the same set of quasinormal modes~\cite{Chandrasekhar:1975abc,Cooper:1994eh} (see also appendix A of~\cite{Berti:2009kk}).

As expected by symmetry, we can write the zero-momentum potentials in this way:
\begin{equation}\begin{split}
(\tilde{V}_S)^{(0)}_{2,2}(r) &= W_{S,2}^2 + \frac{d W_{S,2}}{d r_\star}\, , \\
(\tilde{V}_S)^{(1)}_{2,2}(r) &= W_{S,2}^2 - \frac{d W_{S,2}}{d r_\star} \, , \\
W_{S,2} &= \frac{n}{2} \frac{f S'}{\zeta S} \, .
\end{split}\end{equation}
Curiously, this super potential is exactly equal to the $\mathcal{S}$-deformation we had to do to show stability in the vector sector.

Similarly for the gauge field we can write the potentials as,
\begin{equation}\begin{split}
(\tilde{V}_S)^{(0)}_{1,1}(r) &= W_{S,1}^2 + \frac{d W_{S,1}}{d r_\star}\, , \\
(\tilde{V}_S)^{(1)}_{1,1}(r) &= W_{S,1}^2 - \frac{d W_{S,1}}{d r_\star} \, , \\
W_{S,1} &= -(n-2) \frac{f S'}{2 \zeta S}-\frac{f \phi ' Z'}{2 \zeta  Z} -\frac{f S a'^2 Z}{\zeta  S f'-2 f \zeta  S'}\, .
\end{split}\end{equation}

\section{Scalar sector master scalars}\label{app:scalarsector}
The relation between the gauge invariant fluctuations in the scalar sector and the master scalars is as follows.
For the scalar field and gauge field they are, using $\tilde{k} \equiv \sqrt{k^2 - n K}$ and $\tilde{n} = \sqrt{(n-1)/n}$,
\begin{equation}\label{eq:scalarMasterScalars1}\begin{split}
\text{\boldmath$\varphi$}&= -  \frac{1}{\sqrt{\eta}} \Phi^{(0)}_0 + \frac{k}{n \tilde{n} \tilde{k}} \frac{S \phi'}{S'} \Phi^{(0)}_2 \, , \\
\mathfrak{a}_{r} &= -\frac{\sqrt{2}}{k}  \frac{S \zeta}{f\sqrt{Z}} \partial_t \Phi^{(0)}_1 - \frac{1}{ \tilde{n} k\tilde{k}} \frac{S^2 a'}{f} \partial_t \Phi^{(0)}_2\, , \\
\mathfrak{a}_{t} &=  \frac{1}{ \tilde{n} k \tilde{k}} \frac{S a'}{\mathcal{D} n S'} \left( n S' \left(k^2 S f'+2 f S' \left((n-2) k^2 -n(n-1) K\right)\right) + 2 \zeta^2 k^4 \right) \Phi^{(0)}_2  \\ 
 -& \frac{f }{2 \mathcal{D} \zeta  Z^{3/2}}  \bigg[ \!\!-\!4 \zeta  \sqrt{\eta}  S^2 a' Z^{3/2} \phi' \Phi^{(0)}_0 \!+\! \frac{2 \sqrt{2}}{k}  \mathcal{D} S Z  \partial_r \Phi^{(0)}_1  \\
 +&  \frac{\sqrt{2}}{k}  \left(2 S' Z \left(n S^2 a'^2 Z \!+\! \mathcal{D} (n\!-\!1)\right) \!+\! \mathcal{D} S \phi' Z' \right) \Phi^{(0)}_1 \bigg]   \, .
\end{split}\end{equation}

Note the factors of $1/\tilde{n}$ which, here and below, always occur in front of the gravitational master scalar, indicating the fact that in 3 dimensions there are no dynamical degrees of freedom in the metric. 
For $n = 1$ there is no $h_-$ component and with the same procedure as in appendix \ref{app:specialCasesScalar} above we can show that the same potentials apply to this case.

For the metric fluctuations, we can first express $\mathfrak{h}_{t r}$ directly in terms of the others,
\begin{equation}\label{eq:scalarMasterScalars2}
\mathfrak{h}_{t r} = 2 \partial_t \mathfrak{h}_{r x} + \frac{4 \eta}{k^2} S^2 \phi' \partial_t \text{\boldmath$\varphi$} - \frac{2 n}{k^2} \frac{S S' f}{\zeta^2} \partial_t \mathfrak{h}_{r r} \, .
\end{equation}

The remaining three are as follows,
\begin{equation}\label{eq:scalarMasterScalars3}\begin{split}
\mathfrak{h}_{r r} &= -2    \frac{\zeta^2   S }{\mathcal{D} f} \left(-\sqrt{\eta } \phi ' \mathcal{F} \Phi^{(0)}_0 + \sqrt{2} \zeta  k a' \sqrt{Z(\phi )}  \Phi^{(0)}_1  \right)  + \frac{k}{n \tilde{n}  \tilde{k}} \frac{2  \zeta^2  S}{f  S'} \partial_r \Phi^{(0)}_2 \\
+&\frac{k}{\tilde{n}  \tilde{k}} \frac{\zeta^2}{\mathcal{D} f^2  S'^2} \! \bigg(\! 2  S f' S' \! \left(2 f  S'^2 \!-\! \frac{1}{n}\zeta^2 k^2\right)\!-\! S^2 f'^2 S'^2 \!+\! 4 f \big( \!-\! f  S'^4 \!+\! \zeta^2  S'^2 \!\left(k^2 \!-\! (n \!-\! 1) K\right) \!+\! \frac{\eta}{n^2}  k^2 \zeta^2  S^2 \phi'^2 \\
-&\frac{\eta}{2n} \mathcal{F} S^2 S' \phi'^2 \big) \bigg) \Phi^{(0)}_2   \, ,
\end{split}\end{equation}

\begin{equation}\label{eq:scalarMasterScalars4}\begin{split}
\mathfrak{h}_{r x} &=   \frac{\zeta S^2}{\mathcal{D}} \left( 2 \sqrt{\eta}  \zeta \phi' \Phi^{(0)}_0 - \frac{\sqrt{2}}{k} n S' a' \sqrt{Z} \Phi^{(0)}_1 \right) + \frac{1}{ \tilde{n} k \tilde{k}}S^2 \partial_r \Phi^{(0)}_2 \\
+& \frac{1}{n \tilde{n} k \tilde{k}} \frac{\zeta ^2 S}{  S' f \mathcal{D}} \!\left(\! n S' \!\left(k^2 S f' \!+\! 2 f S' \left((n-2) k^2 \!-\! n(n-1)K\right)\right) \!+\! 2  k^4 \zeta^2 \right)  \Phi^{(0)}_2
\end{split}\end{equation}

\begin{equation}\label{eq:scalarMasterScalars4}\begin{split}
\mathfrak{h}_{t t} &= \frac{f  S }{\sqrt{\eta} \mathcal{D}^2 \zeta \sqrt{Z}} \Bigg( \!\!-\!4\eta   \zeta S f \mathcal{D} \sqrt{Z} \phi ' \partial_r \Phi^{(0)}_0 + \frac{2 \sqrt{2 \eta}}{k} n S S' f \mathcal{D} Z a'  \partial_r \Phi^{(0)}_1 \\
+& \zeta  \sqrt{Z}  \bigg\{ \! \bigg[\! 2 \eta  \!\left(4 f S' \!\left( n (n-2) S'^2 f \!+\! \zeta^2 \!\! \left(k^2\! \!-\! n(n \!-\! 1)K \right)\!\right) \!-\! 2 S f' \!\! \left(\!n(n\!-\!3) S'^2 \!f\! \!+\! k^2 \zeta^2 \right)\!-\!n S^2 f'^2 S'\right)\!\! \bigg] \phi' \\
+& S a'^2 \mathcal{D} Z' \!-\! 2 \zeta ^2 S  \left(- n S' \mathcal{F}+2 \zeta ^2 k^2\right) V' + 4 \eta^2 S^2 f  \mathcal{F} \phi'^3 + 4 \eta n S^2 S' f a'^2 \phi' Z \bigg\}  \Phi^{(0)}_0  \\
-&\frac{\sqrt{2 \eta}}{k} a'\bigg\{ 2 f n^2 S^2 Z^2 a'^2 S'^2+f n S S' Z'  \left(-n S' \mathcal{F}+2 \zeta ^2 k^2\right) \phi' +4 f \zeta ^2 \eta  k^2 S^2 Z \phi'^2  \\
+& 2 Z \!\bigg[ \!2 f^2 (n-1) n^2 S'^4 \!-\! n S f' S' \left(f (n-3) n S'^2+\zeta ^2 k^2\right) \!-\! n^2 S^2 f'^2 S'^2 \!-\! 2 f \zeta ^2 n S'^2 \left(k^2 \!+\! n (n-1) K\right) \!+\! 2 \zeta ^4 k^4\bigg]\\ 
&\bigg\} \Phi^{(0)}_1 \Bigg) - \frac{2}{ \tilde{n} k \tilde{k}} \frac{S^2 f^2}{\zeta^2} \partial_r^2 \Phi^{(0)}_2 - \frac{2}{ \tilde{n} k \tilde{k}} \frac{ f S}{\mathcal{D} \zeta ^3} \bigg[ S f' \left(-f n S \zeta ' S'+f \zeta  (n-2) n S'^2+2 \zeta ^3 k^2\right)+\zeta  n S^2 f'^2 S' \\
-& 2 f \left(f \zeta  n^2 S'^3-f n S \zeta' S'^2+\zeta^3 S' \left(k^2 (1-2 n)+K (n-1) n\right)+\zeta ^2 k^2 S \zeta '\right) \bigg]  \partial_r \Phi^{(0)}_2 \\
+& \frac{1}{n  \tilde{n}  k \tilde{k}} \frac{1}{ S' \mathcal{D}^2} \Bigg\{\!\!-\!k^2 n^2 S^3 f'^3 S'^2 +2 n S^2 f'^2 S' \left(f n S'^2 \left(2 K (n-1) n-3 k^2 (n-2)\right)-2 \zeta^2 k^4\right) \\
+& 4 S f' \!\bigg[\! f n S'^2 \left(f n S'^2 \left(\!-\!(2 n (n-4)\!+\!9)k^2 \!+\! 2n(n-1)(n-2)  K \! \right) \!+\! \zeta ^2 k^2 \left((7 \!-\!3 n)k^2\!+\!n (n \!-\!1) K\right)\right) \!-\! \zeta^4 k^6\bigg] \\
+& 8 f S' \bigg[f n S'^2 \left(f n S'^2 \left( (n (2 n \!-\! 5) \!+\! 4)k^2 \!-\! 2 n (n \!-\!1)^2 K\right) \!+\! \zeta^2 \! \left( \!(\!-((n \!-\! 3) n \!+\! 4))k^4 \!\!+\!\! n (n \!-\!1)\! K k^2 \!\!+\!\! n^2 (n \!-\!1)^2 K^2 \right)\right) \\
+&\zeta ^4 k^4 \left(k^2-n (n-1) K\right) + \frac{n-1}{2} \tilde{k}^2 f S^2 (n^2 Z a'^2 S'^2+2k^2 \eta \zeta^2 \phi'^2 )\bigg] \Bigg\} \Phi^{(0)}_2
\end{split}\end{equation}


\section{Transformations}\label{app:transformations}
\subsection{Regge-Wheeler gauge invariants for the scalar sector}\label{app:transformations:1}
Firstly, let's remind slight difference between using Detweiler gauge and using Detweiler gauge invariants: Detweiler gauge is a certain choice of gauge, where we put all the scalar sector metric coefficients apart from $h_{t t}, h_{t r}, h_{r r}, h_{r x}$ to zero, whereas in Detweiler gauge invariant formulation, we work with $\mathfrak{h}^D_{t t}, \mathfrak{h}^D_{t r}, \mathfrak{h}^D_{r r}, \mathfrak{h}^D_{r x},  \mathfrak{a}^D_{t}, \mathfrak{a}^D_{r}, \text{\boldmath$\varphi$}^D$, which do not feel gauge transformations at all (from now superscripts  $D$ and $RW$ correspond to Detweiler and Regge--Wheeler respectively). Importantly, in Detweiler gauge non-zero quantities correspond exactly to Detweiler gauge invariants: $h_{t t}^D= \mathfrak{h}^D_{tt}, h_{t r}^D=\mathfrak{h}^D_{tr}, h^D_{r r}=\mathfrak{h}^D_{rr}, h^D_{r x}=\mathfrak{h}^D_{rx},  a^D_{t}=\mathfrak{a}^D_{t},   a^D_r=\mathfrak{a}^D_{r}, \varphi^D = \text{\boldmath$\varphi$}^D$ (compare with \ref{eq:gaugeInvariants}), analogously for Regge--Wheeler.

Regge-Wheeler (RW) gauge invariants are another set of gauge invariants which can be build in a way described in \ref{sec:gaugeinv}. In principle, we could build them from the beginning, by ``dressing'' $h_{t t}, h_{t r}, h_{r r}, h_{+}, a_t, a_r, \varphi$ with the remaining metric components to make them gauge invariant. However, RW gauge invariants, as well as any other set of independent gauge invariants in this sector, must be function of Detweiler gauge invariants. To find this relation, we use the previous observation, that in the Detweiler gauge non-zero quantities correspond exactly to Detweiler gauge invariants, the same for RW. It means that it's sufficient to find the transformation between Detweiler and RW gauge and translate it into relation between gauge invariants, which reduces to moving from $h^D_{+}=0$ and $h^D_{rx}\neq0$ to $h^{RW}_{+}\neq0$ and $h^{RW}_{rx}=0$. It can be done by acting with a gauge vector $\zeta^\mu = (0,h^D_{rx},0,..,0)e^{i k x}$. Finally, the relations between Detweiler and RW gauge invariants read:
\begin{equation}\begin{split} 
\mathfrak{h}^{RW}_{tt} =& \mathfrak{h}^{D}_{tt} +\frac{f  f'  }{\zeta  ^2}\mathfrak{h}^D_{rx} \, ,\\
\mathfrak{h}^{RW}_{tr} =& \mathfrak{h}^{D}_{tr}-2 \partial_ t \mathfrak{h}^D_{rx}\, ,\\
\mathfrak{h}^{RW}_{rr} =& \mathfrak{h}^{D}_{rr}+\left(\frac{2 \zeta ' }{\zeta  }-\frac{f' }{f }\right) \mathfrak{h}^D_{rx}-2 \partial_ r \mathfrak{h}^D_{rx}\, ,\\
\mathfrak{h}^{RW}_{+} =& \frac{-2n f  S  S' }{ \zeta  ^2} \mathfrak{h}^D_{rx}\, ,\\
\mathfrak{a}^{RW}_{t} =& \mathfrak{a}^{D}_{t}-\frac{f  a'  }{\zeta  ^2}\mathfrak{h}^D_{rx}\, ,\\
\mathfrak{a}^{RW}_{r} =& \mathfrak{a}^{D}_{r}\, ,\\
\text{\boldmath$\varphi$}^{RW} =&\text{\boldmath$\varphi$}^{D} -\frac{f  \phi '  }{\zeta  ^2}\mathfrak{h}^D_{rx}\, .\\
\end{split}
\end{equation}

\subsection{Transformation between Fefferman-Graham and Eddington-Finkelstein coordinates}\label{app:transformations:2}

Potentials in master equations (\ref{eq:masterequation}) have the same form for both Fefferman-Graham (FG), or Schwarzschild-like, and Eddington-Finkelstein (EF) coordinates. The difference in equations appears only in the form of laplacian - for numerical purposes it's probably more useful to use EF coordinates, since master equations involve only first time derivatives then.

How to express Detweiler gauge invariants in EF coordinates in terms of Detweiler gauge invariants in FG coordinates? (Again, these are distinct quantities related by some functions). 
Let's start with linear metric and gauge vector perturbations, which transform as:
\begin{equation}\label{eq:htransf}\begin{split}
h^{EF}_{\mu\nu} &= L^\alpha_\mu L^\beta_\nu h^{FG}_{\alpha\beta} \, ,\\
a^{EF}_{\mu} &= L^\alpha_\mu a^{FG}_{\alpha} \, ,\\
\end{split}
\end{equation}
where
\begin{equation}
\left (L^\mu_\nu\right) = \begin{pmatrix}
1&-\frac{\zeta}{f}&0\\
0&1&0\\
0&0&\mathbb{1}_n
\end{pmatrix}\, .
\end{equation}
Since we already know how to express $h_{\mu\nu}$ and $a_\mu$ by Detweiler gauge invariants, we can use the transformation rule (\ref{eq:htransf}) to express $\mathfrak{h}^{D,EF}_{\mu\nu}$ and $\mathfrak{a}^{D,EF}_{\mu}$ by $\mathfrak{h}^{D,FG}_{\mu\nu}$ and $\mathfrak{a}^{D,FG}_{\mu}$. For vector sector it reads (we need to add another superscripts: $FG$ or $EF$ and in vector sector we can omit $D$ and $RW$, since they are the same):
\begin{equation}\begin{split}
\mathfrak{h}^{EF}_{tz}& = \mathfrak{h}^{FG}_{tz}\, ,\\
\mathfrak{h}^{EF}_{rz} &= \mathfrak{h}^{FG}_{rz} -\frac{\zeta}{f}  \mathfrak{h}^{FG}_{tz} \, ,\\
\mathfrak{a}^{EF}_{z}& = \mathfrak{a}^{FG}_{z} \, .\\
\end{split}
\end{equation}
and for scalar sector:
\begin{equation}\label{eq:EFFG}
\begin{split}
\mathfrak{h}^{D,EF}_{tt} &= \mathfrak{h}^{D,FG}_{tt}\, ,\\
\mathfrak{h}^{D,EF}_{tr} &= \mathfrak{h}^{D,FG}_{tr} - 2\frac{\zeta}{f} \mathfrak{h}^{D,FG}_{tt}\, ,\\
\mathfrak{h}^{D,EF}_{rr} &= \mathfrak{h}^{D,FG}_{rr}- \frac{\zeta}{f} \mathfrak{h}^{D,FG}_{tr}+\left(\frac{\zeta}{f}\right)^2 \mathfrak{h}^{D,FG}_{tt}\, ,\\
\mathfrak{h}^{D,EF}_{rx} &= \mathfrak{h}^{D,FG}_{rx}\, ,\\
\mathfrak{a}^{D,EF}_{t} &= \mathfrak{a}^{D,FG}_{t}\, ,\\
\mathfrak{a}^{D,EF}_{r} &= \mathfrak{a}^{D,FG}_{r}- \frac{\zeta}{f}\mathfrak{a}^{D,FG}_{t}\, .\\
\end{split}
\end{equation}
RW gauge invariants transform analogously as (\ref{eq:EFFG}), with one difference: $\mathfrak{h}^{RW,EF}_{+} = \mathfrak{h}^{RW,FG}_{+}$.

To fully move to EF coordinates, namely to express $\mathfrak{h}^{D,EF}_{\mu\nu}$ in terms of master scalars like we did for FG coordinates (\eqref{eq:scalarMasterScalars1}-\eqref{eq:scalarMasterScalars4}), one should transform derivatives as well: $\partial_t\rightarrow\partial_t$, $\partial_r \rightarrow \partial_r+\frac{\zeta}{f}\partial_t$. For example, gauge invariants in EF coordinates in vector sector, express by master scalars in the following way:
\begin{equation}\label{eq:vector_final_EF}
\begin{split}
\mathfrak{h}^{EF}_{t z} &\equiv n \frac{f S S'}{\zeta} \Phi^{(1)}_2 + \frac{f S^2}{\zeta} \partial_r \Phi^{(1)}_2 \, , \\
\mathfrak{h}^{EF}_{t z} &\equiv \frac{n f   S   S'  \Phi^{(1)}_2}{\zeta   }+\frac{f   S  ^2 \partial_r \Phi^{(1)}_2 }{\zeta   }+S  ^2 \partial_t \Phi^{(1)}_2 \, , \\
\mathfrak{h}^{EF}_{r z} &\equiv -S^2 \partial_r  \Phi^{(1)}_2-n S S'  \Phi^{(1)}_2\\
\mathfrak{a}^{EF}_{z} &\equiv \sqrt{k^2 - n K} \frac{S}{\sqrt{Z}}\Phi^{(1)}_1 \, .
\end{split}\end{equation}

Having transformation rules from this paragraph and from (\ref{app:transformations:1}) one can move from (\ref{eq:InvsToMasterH1},  \eqref{eq:scalarMasterScalars1}-\eqref{eq:scalarMasterScalars4}) to the desired gauge and coordinate system without performing calculations from the beginning.

\section{Quasinormal modes of AdS planar black holes}\label{app:KS}
In this appendix we derive master scalar wave equations for gravitational  black brane perturbations in our approach. To ease the comparison with the Kovtun-Starinets (KS) approach \cite{Kovtun:2005ev}, widely used in holography, we stick in this section to KS notation and discuss in detail the scalar (sound in KS terminology) sector of perturbations. 
The background line element, eq. (KS-4.2),  reads
\begin{equation}
\label{eq:KSbackground}
ds^2 = a(u) \left( -f(u) dt^2 + dx^2 + dy^2 + dz^2\right) + b(u) du^2 \, ,
\end{equation}    
with
\begin{equation}
a(u) = \frac {\left(r_0/R\right)^2} {u} \qquad \text{and} \qquad 
b(u) = \frac{R^2}{4 u^2 f(u)} \, ,
\end{equation}    
where $u$ is the AdS bulk variable (with the planar black hole horizon located at $u=1$ and AdS boundary at $u=0$), $f(u)=1-u^2$, $R$ is the AdS radius and $r_0$ is related to black hole temperature: $T=r_0/(\pi R^2)$. We take gravitational fluctuations in the form $h_{\mu\nu}(t,u,z)=h_{\mu\nu}(t,u) e^{i q z}$. Under a gauge transformation induced by a gauge vector $\xi_{\mu}(t,u,z) = \xi_{\mu}(t,u) e^{i q z}$ these fluctuations transform as 
\begin{equation}\label{eq:KS_gauge_transformation}
h_{\mu\nu} \rightarrow h_{\mu\nu} - \nabla_{\mu} \xi_{\nu} - \nabla_{\nu} \xi_{\mu} \, .
\end{equation} 
There are seven components of $h_{\mu\nu}$ that enter linearized Einstein equations in the KS sound sector, namely $h_{tt}$, $h_{tu}$, $h_{uu}$, $h_{uz}$, $h_{tz}$, $h_{zz}$, and $h=h_{xx}+h_{yy}$ and out of them four gauge invariant characteristics of fluctuations can be constructed. The Detweiler gauge invariants (i.e. gauge invariants obtained by dressing $h_{tt}$, $h_{tu}$, $h_{uu}$, and $h_{uz}$ with linear combinations of $h_{tz}$, $h_{zz}$, and $h$ and their derivatives) read:
\begin{align}
\mathfrak{h}_{tt} &= h_{tt} + \frac{1}{2} \left( f(u) + \frac{a(u)}{a'(u)} f'(u) \right) h + \frac{2i}{q} \dot{h}_{tz} + \frac{\ddot{h}_{tz}}{2 q^2} - \frac{\ddot{h}_{zz}}{q^2} \,, 
\label{eq:ghtt}
\\
\mathfrak{h}_{tu} & = h_{tu} - \frac{i}{q} \left( \frac{a'(u)}{a(u)} +\frac{f'(u)}{f(u)} \right)  h_{tz} + \frac{i}{q} h'_{tz} - \left(\frac{b(u)}{2 a'(u)} + \frac{1}{4 q^2}\left( \frac{a'(u)}{a(u)} +\frac{f'(u)}{f(u)} \right) \right) \dot h 
\nonumber\\
&+ \frac{1}{2 q^2}\left( \frac{a'(u)}{a(u)} +\frac{f'(u)}{f(u)} \right) \dot h_{zz} + \frac{1}{4 q^2} \dot{h}' - \frac{1}{2 q^2} \dot{h}'_{zz} \,,
\label{eq:ghtu}
\\
\mathfrak{h}_{uu}&=h_{uu} - \frac{b(u)}{a'(u)} h' - \frac{a'(u)b'(u) - 2 b(u) a''(u)}{2 \left(a'(u)\right)^2} h \,,
\label{eq:ghuu}
\\
\mathfrak{h}_{uz}&= h_{uz} - \frac{i a'(u)}{2 q a(u)} h_{zz} + \left( \frac{q^2 b(u)}{2 a'(u)} - \frac{a'(u)}{4 a(u)}\right) h + \frac{h'}{4} - \frac{h'_{zz}}{2} \,,
\label{eq:ghuz}
\end{align}
(and $\mathfrak{h}_{tt}$ corresponds to KS $Z_2$, cf. eq. (KS-3.12)). 
Indeed, it can be easily checked that the above expressions are gauge invariant, and moreover when (\ref{eq:ghtt}-\ref{eq:ghuz}) are solved for $h_{tt}$, $h_{uu}$, $h_{tu}$, and $h_{uz}$ and these solutions are substituted into Einstein equations $E_{\mu\nu}:=R_{\mu\nu}  + \frac{4}{R} g_{\mu\nu}=0$, all gauge dependent terms drop out at linear order and the linearized equations read:
\begin{align}
E_{tu} &= 
\frac{3 u \left( u^2 - 1 \right)}{R^2} \dot{\mathfrak{h}}_{uu} 
+ \frac{q^2 R^2 u}{2 r_0^2} \mathfrak{h}_{tu}
+ \frac{i q R^2 u}{2 r_0^2} \dot{\mathfrak{h}}_{uz} 
+ \mathcal{O}\left( h_{\mu\nu}^2\right)= 0  \,,
\label{eq:Etu}
\\
E_{tz} &= 
  2 i q \left(u^2-1\right) u^2 \dot{\mathfrak{h}}_{uu}
- 2 i q \left(u^2-1\right) u^2 \mathfrak{h}'_{tu}
- 4 i q u^3 \mathfrak{h}_{tu}
- 2 \left(u^2-1\right) u^2 \dot{\mathfrak{h}}'_{uz}
\nonumber\\
& 
+ \mathcal{O}\left( h_{\mu\nu}^2\right)= 0  \,,
\\
E_{uu} &= 
- \frac{R^2 u}{2 r_0^2 \left(u^2-1\right)} \ddot{\mathfrak{h}}_{uu}
+ \frac{2 u \left(u^2-2\right)}{R^2} \mathfrak{h}'_{uu}
+ \left( \frac{q^2 R^2 u}{2 r_0^2} + \frac{4 \left(2 u^4 - 4 u^2 + 1\right)}{R^2 \left(u^2-1\right)} \right) \mathfrak{h}_{uu}
\nonumber\\
& 
+ \frac{R^2 \left(2 u^2-1\right)}{r_0^2 \left(u^2-1\right)^2} \dot{\mathfrak{h}}_{tu}
+ \frac{R^2 u}{r_0^2 \left(u^2-1\right)} \dot{\mathfrak{h}}'_{tu}
- \frac{R^2 u}{2 r_0^2 \left(u^2-1\right)} \mathfrak{h}''_{tt}
- \frac{R^2 \left(u^2-2\right)}{2 r_0^2 \left(u^2-1\right)^2} \mathfrak{h}'_{tt}
\nonumber\\
& 
+ \frac{R^2 u \left(u^2-3\right)}{2 r_0^2 \left(u^2-1\right)^3} \mathfrak{h}_{tt}
+ \frac{i q R^2 u}{r_0^2} \mathfrak{h}'_{uz}
+ \frac{i q R^2 \left(2 u^2-1\right)}{r_0^2 \left(u^2-1\right)} \mathfrak{h}_{uz}
+ \mathcal{O}\left( h_{\mu\nu}^2\right)= 0  \,,
\\
E_{uz} &= 
   \frac{i q u \left(u^2-3\right)}{R^2} \mathfrak{h}_{uu}
+ \frac{i q R^2 u}{2 r_0^2 \left(u^2-1\right)} \dot{\mathfrak{h}}_{tu}
- \frac{i q R^2 u}{2 r_0^2 \left(u^2-1\right)} \mathfrak{h}'_{tt}
+ \frac{i q R^2}{2 r_0^2 \left(u^2-1\right)^2} \mathfrak{h}_{tt}
\nonumber\\
& 
- \frac{R^2 u}{2 r_0^2 \left(u^2-1\right)} \ddot{\mathfrak{h}}_{uz}
+ \mathcal{O}\left( h_{\mu\nu}^2\right)= 0 \,,
\end{align}
\begin{align}
R^2 E_{tt} &= 
\frac{4 r_0^2 u^2 \left(u^2-1\right)^2 \left(u^2+1\right)}{R^4}  \mathfrak{h}_{uu}' 
+ 2 u^2 \left(u^2-1\right) \ddot{\mathfrak{h}}_{uu} 
+ \frac{8 r_0^2 u \left(u^2-1\right) \left(2 u^4-u^2+1\right)}{R^4} \mathfrak{h}_{uu}
\nonumber\\
& 
- u \left(u^2+1\right) \mathfrak{h}_{tt}'
+ 2 u^2 \left(u^2-1\right) \mathfrak{h}_{tt}''
+ \left(\frac{q^2 R^4 u}{2 r_0^2} + \frac{u^4+6 u^2-3}{u^2-1}\right) \mathfrak{h}_{tt}
\nonumber\\
& 
- 2 u \left(u^2+1\right) \dot{\mathfrak{h}}_{tu}
- 4 u^2 \left(u^2-1\right) \dot{\mathfrak{h}}_{tu}'
-2 i q u \left(u^4-1\right) \mathfrak{h}_{zu} + \mathcal{O}\left( h_{\mu\nu}^2\right)= 0 \,,
\\
R^2 E_{zz} &= 
- \frac{4 r_0^2 \left(u^2-1\right)^2 u^2}{R^4} \mathfrak{h}'_{uu}
- \frac{2 \left(u^2-1\right) u \left(q^2 R^4 u+r_0^2 \left(8 u^2+4\right)\right)}{R^4} \mathfrak{h}_{uu}
+ 2 u \dot{\mathfrak{h}}_{tu}
\nonumber\\
& 
- u \mathfrak{h}'_{tt}
+ \frac{ \left(q^2 R^4 u+2 r_0^2 \left(u^2+1\right)\right)}{2 r_0^2 \left(u^2-1\right)} \mathfrak{h}_{tt}
- 4 i q \left(u^2-1\right) u^2 \mathfrak{h}'_{uz}
- 2 i q \left(3 u^2+1\right) u \mathfrak{h}_{uz}
\nonumber\\
& 
+ \mathcal{O}\left( h_{\mu\nu}^2\right)= 0  \,,
\end{align}
\begin{align}
R^2 \left(E_{xx} + E_{yy}\right) &= 
- \frac{4 r_0^2 u^2 \left(u^2-1\right)^2 \mathfrak{h}'_{uu}}{R^4} 
+ \frac{8 r_0^2 u \left(-2 u^4+u^2+1\right)}{R^4} \mathfrak{h}_{uu}
+ 2 u \dot{\mathfrak{h}}_{tu} 
\nonumber\\
&
- u \mathfrak{h}'_{tt}
+ \frac{\left(u^2+1\right)}{u^2-1} \mathfrak{h}_{tt}
+ 2 i q u \left(u^2-1\right) \mathfrak{h}_{uz}
+ \mathcal{O}\left( h_{\mu\nu}^2\right)= 0  \,.
\label{eq:E+}
\end{align}

Now, we make our ansatz that the gauge invariant characteristics of perturbations are given in terms of linear combinations of a single master scalar $\Phi(t,u)$ satisfying scalar wave equation on the background solution (\ref{eq:KSbackground}), namely
\begin{equation}
\label{eq:BBmasterScalar}
(-\Box  + V) \Phi(t,u) e^{i q z} = 0 \,,
\end{equation}
where $\Box$ is the scalar wave operator corresponding to line element (\ref{eq:KSbackground}). Plugging such ansatz into linearized Einstein equations (\ref{eq:Etu}-\ref{eq:E+}) one gets uniquely defined formulas, namely 
\begin{align}
\mathfrak{h}_{tt} &= \frac{8 \left(r_0/R\right)^2}{3 u^2 \left(2 \mathfrak{q}^2+3 u\right)^2} \left[2 \mathfrak{q}^2 \left(6 \mathfrak{q}^2 f(u)+6 \mathfrak{q}^2+9 u^5+6
   \mathfrak{q}^2 u^4+\left(4 \mathfrak{q}^4-27\right) u^3+27 u\right) \Phi(t,u) \right.
\nonumber\\
& \left. -3 u f(u) \left(2 \mathfrak{q}^2+3 u\right) \left(\left(4 \mathfrak{q}^2 + 3 u^3 + 3u \right) \partial_u \Phi(t,u) - u f(u) \left(2 \mathfrak{q}^2+3 u\right) \partial_{u u} \Phi(t,u)\right)\right]
\\
\mathfrak{h}_{tu} &= R^2 \frac{3 u f(u) \left(2 \mathfrak{q}^2 + 3 u\right) \partial_{t u} \Phi(t,u) + 2 \left(-12 \mathfrak{q}^2 f(u) + 9 \mathfrak{q}^2 + 9 u^3 + 2 \mathfrak{q}^4 u\right) \partial_t \Phi(t,u)}{3 u^2 f(u) \left(2 \mathfrak{q}^2 + 3 u\right)}
\\
\mathfrak{h}_{zu} &= 2 i r_0 \mathfrak{q} \left(\frac{2 \mathfrak{q}^2 \left(2 \mathfrak{q}^2 u + 3\right)}{3 u f(u) \left(2 \mathfrak{q}^2+3 u\right)} \Phi(t,u) - \frac{1}{u} \partial_u \Phi(t,u)\right)
\\
\mathfrak{h}_{uu} &= R^2 \frac{4 \mathfrak{q}^2 \left[-\left(4 \mathfrak{q}^2 f(u) - 2 \mathfrak{q}^2 - 3 u^3\right) \Phi(t,u) + u f(u) \left(2 \mathfrak{q}^2 + 3 u\right) \partial_u \Phi(t,u) \right]}{3 u^2 f^2(u) \left(2 \mathfrak{q}^2 + 3 u\right)}
\end{align}
and 
\begin{equation}
\label{eq:BBpotential}
V = -\frac{8 \mathfrak{q}^2 \left(2 \mathfrak{q}^2 + 3 u^3 + 6\mathfrak{q}^2 u^2 + 9 u \right)}{R^2 \left(2 \mathfrak{q}^2 + 3 u \right)^2} \, ,
\end{equation}

where $\mathfrak{q} = q/(2\pi T) = q R^2/(2r_0)$, cf. (KS-4.6). Now, from the master scalar wave equation (\ref{eq:BBmasterScalar}) quasinormal modes of the planar black hole can be effectively computed with \cite{Jansen:2017oag} algorithm. 
The key point to be noted is that the scalar wave equation (\ref{eq:BBmasterScalar}) is by definition \textit{covariant} i.e. once the form of the potential is found in one coordinate system, for example (\ref{eq:KSbackground}) used here, it can be easily transformed (as a scalar) to any other coordinate system. 

Now we compare the KS and KI approaches for doing numerics, where we focus on the approach of discretizing the QNM equation on a pseudospectral grid and computing the generalized eigenvalues, an approach first used in gravity in \cite{Dias:2009iu}, and which is used in the publicly available package \textit{QNMspectral}\cite{Jansen:2017oag}.

First, the only time derivatives come from the Laplacian, which in Eddington-Finkelstein coordinates gives a linear dependence of the equations on the quasinormal mode frequency. 
This is of great practical convenience, since it naturally has the form of a generalized eigenvalue equation when discretized.
In contrast, the KS approach introduces higher orders of the frequency into the equation through the decoupling process, up to fourth order in the sound channel. 
In order to turn this equation into a generalized eigenvalue equation one has to linearize it in the frequency by introducing extra functions $\omega^p \Phi$ and writing it as a coupled system of equations, effectively increasing the size of the matrix by a factor of four.

Second, from the KS equation one obtains a lot of numerical artifacts near $\omega = \pm k$, which are approximate solutions of the discretized equation but not physical solutions. In the KI form these are completely absent.

Finally, at the same level of numerical accuracy the KI equations give much more accurate results for the physical frequencies, even when disregarding the factor 4 increase in matrix size due to the higher powers of the frequency.

As a quantitative illustration we compute the QNMs at $\mathfrak{q} = 1$ using both equations, with the package \cite{Jansen:2017oag}. 
From the KS approach we use (KS-4.35) with the replacement $Z_2(u) = u^2 (1-u)^{- i \omega/2} \tilde{Z_2}(u)$, so that the normalizable and ingoing mode is regular at both endpoints.
From the KI approach, we use Eq. (\ref{eq:BBmasterScalar}) but in Eddington-Finkelstein coordinates $ds^2 = - f(u) dt^2 - 2 u^{-2} du dt + u^{-1}(dx^2 + dy^2 + dz^2)$, with $f(u) = u^{-2} (1 - u^4)$. The potential is simply Eq. (\ref{eq:BBpotential}) with the replacement $u \rightarrow u^2$.
We also have to rescale the master scalar as $\Phi = u^2 \tilde{\Phi}$, where now the non-normalizable term goes as $\tilde{\Phi} \sim \log(u)$ and the normalizable term goes to a constant.

For both we use $N = 40$ grid points and check for convergence by comparing with the same computation at $N=50$ grid points.
so we are comparing the eigenvalues of a $40 \times 40$ matrix to a $40^4 \times 40^4$ matrix.
Illustrating the points made above, the KS computation takes about 150 ms versus about 10 ms for the KI computation.
The KS computation has many unphysical poles around $\mathfrak{w} = \pm 1$ while the KI has none.
Finally we show the physical modes that are visible with this accuracy in Table \ref{tab:KScomparison}.

\begin{table}[htp]
\begin{center}
\begin{tabular}{c||c |}
j & $\mathfrak{w}_j$  \\ \hline\hline
0 & $\pm \underline{0.741429965}5 - \underline{0.286280007}2 i $ \\ \hline
1 & $\pm \underline{1.73351}1095 - \underline{1.34300}7549 i $ \\ \hline
2 & $\pm \underline{2.70}5540 - \underline{2.35}7062 i $ \\ \hline
3 & $\pm 3.689 - 3.364 i $ \\ \hline
4 & $\pm 4.7 - 4.4 i $  \\ \hline
\end{tabular}
\end{center}

\caption{QNMs in sound channel of AdS$_5$-Schwarzschild black brane at momentum $\mathfrak{q} = 1$. 
Full results are computed using the KI approach, only showing converged digits.
Underlined digits are what is visible at the same numerical accuracy using the KS equation.}
\label{tab:KScomparison}
\end{table}%

A final curiosity is that as one takes the zero-momentum limit in the planar case, one obtains different potentials for different helicities, even though without any momentum these can no longer be distinguished.
 On the other hand in the Kovtun-Starinets formalism one obtains identical equations independent of the helicity, only of the type of field (metric, gauge field or scalar). 
 It turns out however that these different potentials actually give rise to the same spectrum of QNMs (as tested in Reissner-Nordstr\"{o}m backgrounds in various dimensions), so the potentials are iso-spectral.

\end{document}